\documentclass[twocolumn,floatfix,amssymb,superscriptaddress,showpacs]{revtex4}
\usepackage{amssymb}
\usepackage{amsmath}
\usepackage{amsfonts}
\usepackage{graphicx}
\usepackage{color}
\usepackage{bm}

\renewcommand{\Re}{\,\textrm{Re}\,}

\DeclareMathOperator{\Tr}{Tr}
 
\DeclareMathOperator{\sgn}{sgn}
\DeclareMathOperator{\sh}{sh}
\DeclareMathOperator{\ch}{ch}
\newcommand{\bb}{\begin{equation}}
\newcommand{\ee}{\end{equation}}

\begin{document}

\title{Co-tunneling current through the two-level quantum dot coupled to magnetic leads: A role of exchange interaction}
\author{A.U.\,Sharafutdinov \email{e-mail: shazat@itp.ac.ru}}
\affiliation{Moscow Institute of Physics and Technology, 141700 Moscow, Russia}

\author{I.S.\,Burmistrov}
\affiliation{L.D. Landau Institute for Theoretical Physics, Kosygina
  street 2, 117940 Moscow, Russia}

\begin{abstract}
 The co-tunneling current through a two-level doubly occupied quantum dot weakly coupled to  ferromagnetic leads is calculated in the Coulomb blockade regime. It is shown that the dependence of the differrential conductance on applied voltage has a stair-case structure with different sets of ``stairs" for parallel and anti-parallel configurations of magnetization of the leads.  Contributions to the current from elastic and inelastic processes are considered distinctly. It is observed that the interference part of the co-tunneling current involves terms corresponding to inelastic processes.  Dependence of the co-tunneling current on the phases of the tunneling amplitudes is studied. 
\end{abstract}


\maketitle

\section{Introduction}

It is well-known that the resistivity of a conductor depends on its magnetic state~\cite{Abrikosov}.
There are several quantum effects increasing this dependence, e.g. Kondo effect ~\cite{Kondo}, colossal magnetoresistance effect~\cite{colMag}, giant magnetoresistance effect ~\cite{GiantMag}. 
One can observe giant magnetoresistance effect in thin film structures composed of alternating ferromagnetic and non-magnetic layers.  At the room temperature resistance for the parallel configuration of magnetizations in the leads is a few percent less than the resistance for the antiparallel configuration. Thus spin degrees of freedom can considerably influence the magnetoresistance of the system. 

Recently this influence of magnetism on transport has attracted a lot of attention in different systems, in particular, in quantum dots~\cite{QD_REF1,QD_REF2,QD_REF3,Kurland,ABG,QD_REF4,QD_REF5,QD_REF6,QD_REF7,QD_REF8,QD_REF8a}.
A quantum dot is the simplest system in which one can study interplay between spin and charge degrees of freedom. If the size of the dot is small enough the characteristic electrostatic energy which is necessary to change the number of electrons on the dot can exceed characteristic temperature and applied source-drain voltage ($V$). Such regime of electron transport is referred to as Coulomb blockade. In this regime processes of the first order in tunneling amplitudes (sequential tunneling) are not suppressed only at specific values of the gate voltage (so-called Coulomb peaks) at which the charging levels for electrons on the dot become degenerate in the absence of tunneling~\cite{KulikShekhter}. Therefore the current outside the Coulomb peaks is determined by co-tunneling processes. They are of the second order:  tunneling of electrons from one lead to another through a virtual state on the dot~\cite{NazarovAverinOdintsov}.

Addional factor which strongly effects current through a quantum dot at low temperatures ($T$) is discreteness of the energy spectrum~\cite{QDRev}. 
For example, in the limit $T\gg \Delta$ the conductance due to inelastic co-tunneling is proportional to $(e^2/h)(T/E_c)^2$ ~\cite{NazarovAverinOdintsov} where $\Delta$ denotes the single-particle spacing, $e$ the electron charge, $h$ the Planck's constant, and $E_c = e^2/(2C)$ the electrostatic energy of the quantum dot with the total capacitance $C$. In the opposite regime of low temperatures $T\ll \Delta$ the (inelastic co-tunneling) conductance is proportional to $(e^2/h) (\Delta^3/E_c^2T)\exp(-\Delta/T)$, i.e. exponentially suppressed  with temperature~\cite{KangMin}. The same holds for the conductance due to elastic co-tunneling: at high temperatures $T\gg \Delta$ the conductance is $\propto (e^2/h)\Delta/E_c$~\cite{NazarovAverinOdintsov} whereas for $T\ll \Delta$ it is proportional to $(e^2/h)(\Delta/E_c)^2 $~\cite{KangMin}. We mention that at low temperatures the elastic co-tunneling contribution dominates the inelastic one.

The presence of exchange interaction of electrons on the quantum dot results in dependence of the conductance on magnetic polarization of the leads. 
Recently this dependence has been extensively studied in the two-level quantum dot~\cite{BarnasWeymann}. In the paper~\cite{Weymann} 
numerical solution of the rate equations describing 
transport through the two-level quantum dot was employed. 

Contrary to papers~\cite{BarnasWeymann,Weymann}, our aim is to calculate analytically co-tunneling current through the two-level quantum dot with 
exchange interaction coupled to magnetic leads. Then we examine interference and non-interference contributions to the co-tunneling current and contributions corresponding to elastic and inelastic processes. 
We obtain that at non-zero exchange energy ($J>0$) the interference part of the co-tunneling current for antiparallel alignment of magnetization in the leads contains inelastic terms which correspond to transitions between states of a quantum dot with different values of the total spin. This fact differs problem under consideration from the standard one~\cite{NazarovAverinOdintsov}. Also we demonstrate that low temperature asymptotics of the current are similar to that of in the two-level quantum dot without exchange interaction ~\cite{KangMin} if we substitute $|\Delta-2J|$ for the level spacing $\Delta$.
At $J=\Delta/2$ when the transition from the singlet ($J<\Delta/2$) to triplet ($J>\Delta/2$) ground state occurs, some of inelastic processes become elastic. In addition we find that low-temperature current-voltage characteristics are non-linear. However, they became linear if the spin-triplet gap $|\Delta-2J|$ is equal to zero. This fact can be used to observe the spin-triplet transition point experimentally.  We have found that inelastic co-tunneling current  has  singularities corresponding to transitions between energy levels of the quantum dot.

In general, contributions to the conductance of higher orders (4th and higher) in the tunneling amplitudes contain logarithmic divergences~\cite{GlazmanPustilnik}. These divergences ~\cite{Martinek} are not relevant if $e|V|,T\gg T_K \sim \Delta \exp(-\Delta/\nu |t|^2)$ where $\nu$ and $t$ stand for the density of states in the leads and the tunneling amplitude, respectively. Thus the co-tunneling processes provide the main contribution to the current.

To observe the interplay of ferromagnetism and discretness of energy spectrum in quantum dots one should prepare a quantum dot with large level spacing. In ultra-small metallic islands (nanoparticles)~\cite{Al} or in semiconducting quantum dots ~\cite{SC}, the level spacing is comparable with the charging energy or even larger. This also happens in the case
of a molecule attached to metallic leads~\cite{Molecule} and in two-dimensional electron gas with ferromagnetic leads~\cite{Hamaya}.

The paper is organized as follows. We start with the formulation of the model describing the system under consideration (Sec.~\ref{Sec_Model}). 
Then, in Sec.~\ref{Sec_Current}, we derive general expression for the co-tunneling current under several approximations which do not change qualitative properties of the system. 
In Section~\ref{Sec_ccurrent}  we present the expressions for the non-interference and interference contributions to the co-tunneling current. Finally, we present discussions and conclusions in Sec.~\ref{Sec_Conc}.

\section{\label{Sec_Model} The Model}

The quantum dot coupled to the leads is described by the following Hamiltonian:
\bb
H=H_l+H_r+H_{QD}+H_T.
\ee
Here $H_{j}=\sum \epsilon_{j\sigma}a^{\dag}_{j\sigma}a_{j\sigma}$, 
 $j=l,r$,  are Hamiltonians of the left and right leads correspondingly with $a_{jk\sigma}/a^{\dag}_{jk\sigma}$ denoting annihilation/creation operators in the corresponding lead, $\sigma$ - spin index. Term
\bb H_T=\sum_{j=l,r} \sum_{k,\alpha,\sigma,\sigma^\prime} t^{(j)}_{k\alpha,\sigma
\sigma^\prime}a^{\dag}_{jk\sigma}d_{\alpha\sigma^\prime}+h.c. 
\ee
describes the tunneling between the quantum dot and lead where $d_{\alpha\sigma}$ stands for an annihilation operator of electrons on the dot. The quantum dot is modeled by the so-called universal Hamiltonian ~\cite{Kurland}
\bb H_{QD}=\sum_{\alpha,\sigma}\epsilon_{\alpha\sigma} d^{\dag}_{\alpha\sigma}d_{\alpha\sigma}+E_c(\hat{N}-N_0)^2-J\hat{\bm{S}}^2 ,\label{Hqd}
\ee
where $\hat{N}=\sum_{\epsilon\alpha}d^{\dag}_{\epsilon\alpha}d_{\epsilon\alpha}$ denotes the operator of the number of particles on the dot, $\hat{\bm{S}}=\sum_{\sigma\sigma^\prime} d^\dag_{\alpha\sigma}\bm{\sigma}_{\sigma\sigma^\prime}d_{\alpha\sigma^\prime}$  the  operator of the total spin of electrons, $N_0$ the equlibrium number of electrons on the dot which minimizes the electrostatic energy and is  tunable by the gate voltage, and $J>0$ the exchange energy. 
\begin{figure}[t]
 \includegraphics[width=8cm,keepaspectratio]{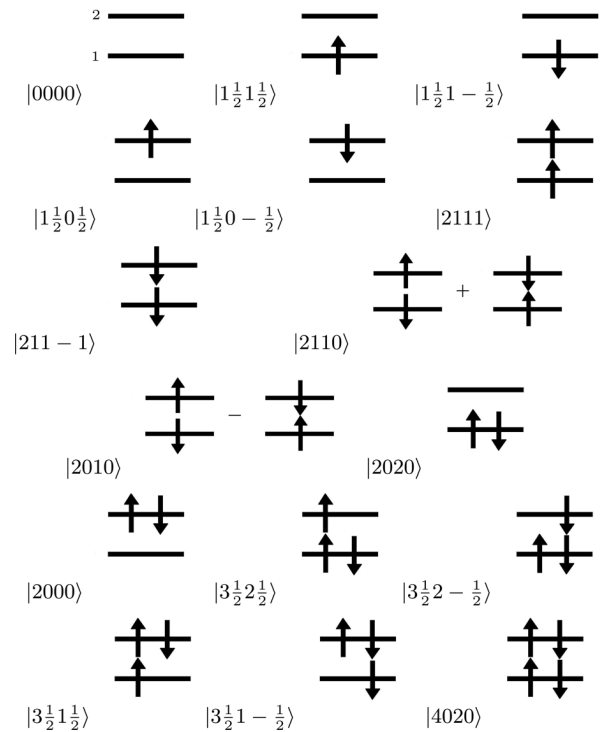}
 \caption{Eigen states of the isolated two-level quantum dot.}
 \label{FigureLevels}
\end{figure}

The main objective of the present paper is analytical investigation of interplay of spin and charge degrees of freedom in transport in the co-tunneling regime. The simplest system revealing such effects is the quantum dot with two single-particle levels. Although Hamiltonian \eqref{Hqd} is derived under assumption that the number of levels on the dot involved in the transport is large~\cite{Kurland}, it is widely used to model a two-level quantum dot~\cite{BarnasWeymann,Weymann,Tarucha,Paaske}. As we demonstrate in Appendix~\ref{App1} Hamiltonian~\eqref{Hqd} can adequately describe the many-particle spectrum even for a quantum dot with only two low-lying single-particle levels. 

In what follows we consider two-level quantum dot, i.e.  $\epsilon_\alpha$ will take only two values
 $\epsilon_1$ and $\epsilon_2=\epsilon_1+\Delta$.
There are  $16$ (many-particle) eigen states. The energy of a state denoted by $|NSnM\rangle$ is given by the following expression  
\begin{equation}
E_{NSnM}=E_c(N-N_0)^2+n\epsilon_1+(N-n)\epsilon_2-JS(S+1) .
\end{equation}
Here $N$  denotes the number of electrons on the dot, $S$ the total spin of the dot, and $n$ the number of electrons on the first level. The eigen states of the isolated quantum dot are depicted on Fig.~\ref{FigureLevels}.

\section{Contribution to the current of the fourth order in $t^{(j)}_{k\alpha,\sigma\sigma^\prime}$\label{Sec_Current}}

\subsection{Perturbation theory}

Let us assume that the Coulomb energy is much larger than other energy scales,  $E_c\gg T,\Delta, J,$ and $N_0$
is close to the integer number (regime of a Coulomb valley). 
Then the main contribution to the current at low temperatures is of the 4th order in $t^{(j)}_{k\alpha,\sigma\sigma^\prime}$  ~\cite{NazarovAverinOdintsov}. Indeed the odd terms vanish. The second order term corresponding to real transitions changing the charge of the dot is exponentially suppressed $(\sim \exp(-E_c/T))$ for almost all gate voltages except those which correspond to Coulomb peaks ~\cite{KulikShekhter}.
The contribution to the current of the fourth order describes transitions of an electron from lead to lead trough the virtual states on the dot. These transitions are referred to as co-tunneling of electrons.

Let us consider the equation for the density matrix in the interaction representation:
\bb\dot{\rho}=-i[H_T,\rho] .\ee
By solving it perturbatively, we find for the third order correction~\cite{Mahan}
\begin{eqnarray}
\rho^{(3)}(t_3)&=&i \int_{-\infty}^{t_3}dt_2\int_{-\infty}^{t_2}dt_1\int_{-\infty}^{t_1}dt_0 \notag \\
&\times & [H_T(t_2), [H_T(t_1),[H_T(t_0),\rho_0]]] 
\end{eqnarray}
where $\rho_0 = \exp(-\beta (H_l+H_r+H_{QD}))$ stands for the equlibrium density matrix of the unperturbed system ($\beta=1/T$). The current operator is equal to derivative of the charge operator in one of the leads:
\begin{equation}
 \hat{I_j}=e\dot{\hat N}_j=
i\hat{X}-i\hat{X}^\dag , \quad
\hat{X}\equiv \sum\limits_{k\alpha\sigma\sigma^\prime} t^l_{k\alpha;\sigma\sigma^\prime}a^\dag_{lk\sigma}d_{\alpha\sigma^\prime}  .
\end{equation}
It is worth to mention that the current is one order higher in tunneling amplitudes than the density matrix. 
The fourth order in  $t^{(j)}_{k\alpha,\sigma\sigma^\prime}$ correction reads
\begin{eqnarray}
I^{(4)}&=& \frac{1}{Z}\Tr(\rho_3 \hat{I_l})\notag\\
&=&
\frac{2e}{Z} \Re \int_{-\infty}^t dt_1\int_{-\infty}^{t_1} dt_2\int_{-\infty}^{t_2} dt_3  \notag\\
&\times & \Bigl(\langle X^\dag(t)H_T(t_1)H_T(t_2)H_T(t_3)\rangle \notag\\
&-&\langle H_T(t_3)X^\dag(t)H_T(t_1)H_T(t_2)\rangle\notag\\&-&\langle H_T(t_2)X^\dag(t)H_T(t_1)H_T(t_3)\rangle\notag\\&+&\langle H_T(t_3)H_T(t_2)X^\dag(t)H_T(t_1)\rangle\notag\\&-&\langle H_T(t_1)X^\dag(t)H_T(t_2)H_T(t_3)\rangle\notag\\&+&\langle H_T(t_3)H_T(t_1)X^\dag(t)H_T(t_2))\rangle\notag\\&+&\langle H_T(t_2)H_T(t_1)X^\dag(t)H_T(t_3)\rangle\notag\\&-&\langle H_T(t_3)H_T(t_2)H_T(t_1)X^\dag(t)\rangle \Bigr ) \label{Current4}
\end{eqnarray}
where $Z\equiv \Tr(\rho_0)$ is the grand canonical partition function and  $\langle\dots\rangle\equiv \Tr(\dots\rho_0)/Z$.

Let us express all of the averages in Eq.~\eqref{Current4} in terms of the  exact two-particle correlators for the isolated dot ($H_{QD}$) and the following Green functions of electrons in the leads:
\begin{gather}
\begin{split}
G^{j>}_{\beta_1\beta_2}(t,t') \equiv -i\langle a_{j\beta_1}(t)a^\dag_{j\beta_2}(t') \rangle,  \\
G^{j<}_{\beta_1\beta_2}(t,t') \equiv\hspace{0.3cm} i\langle a_{j\beta_1}(t)a^\dag_{j\beta_2}(t') \rangle.
\end{split}
\end{gather}
Here we introduce $\beta = \{k\sigma\}$.
Each of the eight terms in Eq.~\eqref{Current4} is evaluated in Appendix (cf. Eqs~\eqref{C1}-\eqref{C8}).
For example,
\begin{eqnarray}
& & \hspace{-0.5cm} \langle X^\dag(t)H_T(t_1)H_T(t_2)H_T(t_3)\rangle \notag
\\
&=&\hspace{-0.1cm}\langle d^\dag_{\alpha_1t}d_{\alpha_2t_1}d^\dag_{\alpha_3t_2}d_{\alpha_4t_3}\rangle\,
\tau_{1432}
G^{l>}_{\beta_1\beta_4}(t,t_3)G^{r<}_{\beta_3\beta_2}(t_2,t_1)
\notag\\
&-&\langle d^\dag_{\alpha_1t}d_{\alpha_2t_1}d^\dag_{\alpha_3t_2}d_{\alpha_4t_3}\rangle\,
\tau_{1234}
G^{l>}_{\beta_1\beta_2}(t,t_1)G^{r>}_{\beta_3\beta_4}(t_2,t_3)
\notag\\
&+&\langle d^\dag_{\alpha_1t}d^\dag_{\alpha_2t_1}d_{\alpha_3t_2}d_{\alpha_4t_3}\rangle\,
\tau_{1324}
G^{l>}_{\beta_1\beta_3}(t,t_2)G^{r>}_{\beta_2\beta_4}(t_1,t_3)
\notag\\
&-&\langle d^\dag_{\alpha_1t}d^\dag_{\alpha_2t_1}d_{\alpha_3t_2}d_{\alpha_4t_3}\rangle\,
\tau_{1423}
G^{l>}_{\beta_1\beta_4}(t,t_3)G^{r>}_{\beta_2\beta_3}(t_1,t_2)
\notag\\
&+&\langle d^\dag_{\alpha_1t}d_{\alpha_2t_1}d_{\alpha_3t_2}d^\dag_{\alpha_4t_3}\rangle\,
\tau_{1243}
G^{l>}_{\beta_1\beta_2}(t,t_1)G^{r<}_{\beta_4\beta_3}(t_3,t_2)
\notag\\
&-&\notag\langle d^\dag_{\alpha_1t}d_{\alpha_2t_1}d_{\alpha_3t_2}d^\dag_{\alpha_4t_3}\rangle\,
\tau_{1342}
G^{l>}_{\beta_1\beta_3}(t,t_2)G^{r<}_{\beta_4\beta_2}(t_3,t_1) , \\
&& \label{MC1}
\end{eqnarray}
where $\langle \dots \rangle = \Tr \dots e^{-\beta H_{QD}}/\Tr e^{-\beta H_{QD}}$, $\alpha_k = \{\alpha,\sigma\}$,
$\tau_{ijkl}=\overline{t}^l_{\beta_i\alpha_i}t^l_{\beta_j\alpha_j}\overline{t}^r_{\beta_k\alpha_k}t^r_{\beta_l\alpha_l}$, and $d_{\alpha t}\equiv d_{\alpha}(t)$.
We mention that Eq.~\eqref{MC1} involves terms which are proportional to: 
$\overline{t}^l_{\beta_1\alpha_1}t^l_{\beta_2\alpha_2}\overline{t}^l_{\beta_3\alpha_3}t^l_{\beta_4\alpha_4}$ and $\overline{t}^r_{\beta_1\alpha_1}t^r_{\beta_2\alpha_2}\overline{t}^r_{\beta_3\alpha_3}t^r_{\beta_4\alpha_4}$.
If we set, e.g. $t^r=0$, the current should vanish. 
However, the terms proportional to
$\overline{t}^l_{\beta_1\alpha_1}t^l_{\beta_2\alpha_2}\overline{t}^l_{\beta_3\alpha_3}t^l_{\beta_4\alpha_4}$
remain unchanged.
Thus such terms give no contribution to the current in the fourth order. Therefore we shall omit them in what follows.

Due to the presence of interactions in $H_{QD}$, correlators of the form $\langle d^\dag d d^\dag d\rangle$ in Eq.~\eqref{MC1} cannot be simplified with the help of the Wick theorem.
In general absence of the Wick theorem leads to a very tedious expression for the current. Therefore we introduce some simplifications which do not affect qualitative properties of the system but allows analytical calculation of correlators  $\langle d^\dag d d^\dag d\rangle$.

\subsection{Approximations}

We calculate the 4th order correction to the current, Eq.~\eqref{Current4} under the following assumptions.
\begin{enumerate}

\item[(i)] The leads are made of ferromagnetic metal with magnetization along some axis $z$.
Considering the exchange interaction to be isotropic we obtain that the Green function of electrons in the lead:
$G^{j>}_{k_1k_2\sigma_1\sigma_2}(t,t^\prime) \equiv  -i\langle a_{jk_1\sigma_1}(t)a^\dag_{jk_2\sigma_2}(t^\prime) \rangle $  is proportional to
$A(k_1,k_2,t,t^\prime)\delta_{\sigma_1,\sigma_2}+B(k_1,k_2,t,t^\prime)\sigma^z_{\sigma_1,\sigma_2}$, i.e. it is diagonal in the spin indices $\sigma_1,\sigma_2$.  
Under this assumptions Green functions read ($j=l,r$)
\begin{eqnarray}
G^{j>}_{k_1k_2\sigma_1\sigma_2}(t,t^\prime) &=&-i\delta_{k_1k_2}\delta_{\sigma_1\sigma_2}\mathcal{Z}^{(j)}_{\sigma_1}\int\frac{d\varepsilon}{2\pi}(1-n^{(j)}_F(\varepsilon)) \notag \\ &\times& e^{-i\varepsilon(t-t^\prime) }\delta(\varepsilon-\epsilon_{k_1\sigma_1})\notag \\
 &\equiv&  -i\delta_{k_1k_2}\delta_{\sigma_1\sigma_2}\mathcal{Z}^{(j)}_{\sigma_1}G^{j>}_{k_1\sigma_1}(t,t^\prime) ,
\\
G^{j<}_{k_1k_2\sigma_1\sigma_2}(t,t^\prime) 
&=& i\delta_{k_1k_2}\delta_{\sigma_1\sigma_2}\mathcal{Z}^{(j)}_{\sigma_1}\int\frac{d\varepsilon}{2\pi}n^{(j)}_F(\varepsilon)\notag\\
&\times & e^{-i\varepsilon (t-t^\prime) }\delta(\varepsilon-\epsilon_{k_1\sigma_1})
\notag \\
&\equiv &i\delta_{k_1k_2}\delta_{\sigma_1\sigma_2}\mathcal{Z}^{(j)}_{\sigma_1}G^{j<}_{k_1\sigma_1}(t,t^\prime) .
\end{eqnarray}
Here $\epsilon_{k\sigma}$ stands for the energy of a single-particle excitation in the leads
and coefficient $\mathcal{Z}^{(m)}_{\sigma}$   arises due to renormalization of the spectral density due to interactions in the leads.

\item[(ii)]  Next we assume that there are no magnetic impurities in the tunneling junctions. Therefore we neglect the probability for the spin to flip during the tunneling, i.e.
\bb t^{(l,r)}_{k \alpha\sigma_1\sigma_2}\equiv t^{(l,r)}_{k \alpha \sigma_1}\delta_{\sigma_1\sigma_2} .\ee
Also since only energies near the Fermi level in the leads are essential for the calculation of the current we will ignore the dependence of the tunneling amplitudes on energy and introduce (dimensionless) tunneling conductances
\begin{eqnarray}
g^{j
}_{\alpha\sigma} &=& \frac{1}{\Delta}\sum_k \mathcal{Z}^{(j)}_{\sigma}\delta(E_F-\epsilon_{k\sigma})|t^j_{k\epsilon\sigma}|^2.
\end{eqnarray}
  
  \item[(iii)] Finally, we restrict our consideration to the states with only two electrons on the dot. If  $\Delta>2J$ the spin in the ground state is zero  (both electrons occupy the lowest energy level). In the opposite case of $\Delta<2J$ the ground state is ferromagnetic (total spin is unity) such that each energy level is singly occupied.
Therefore it is possible to observe drastic effect in the transport through the quantum dot due to the change of the spin in the ground state.
In what follows we assume such value of the gate voltage that the only important states of the system are the states with two electrons on the dot, i.e. $N_0 \simeq 2$.   
\end{enumerate}

In the current ~\eqref{Current4} one can distinguish terms of two types.
Terms of the first type depend only on the absolute values of the tunneling amplitudes ($\propto|t^l |^2|t^r|^2$). Contributions of the second type involve also  relative phases of the tunneling amplitudes. 
One can refer to corrections of the first type as non-interference whereas the second type as interference contributions. 
To single out effects associated with dependence of the current on the phases of the tunneling amplitudes it is convenient to analyse interference and non-interference contributions to the current separately.
As well-known~\cite{NazarovAverinOdintsov}, there are processes of two types: inelastic processes during which the energy of the quantum dot changes and elastic with the same energies of the initial and final states of the quantum dot.  
Due to conservation of the energy of the whole system we have: $\epsilon_1 +E_i = \epsilon_2+E_f$, where $\epsilon_1$, $\epsilon_2$ denote energies of an electron before and after the tunneling event, $E_i$, $E_f$ energies of the initial and final states of the dot. By definition inelastic co-tunneling involves change of the energy of the dot:  $E_i\neq E_f$, (electron-hole pair arises).
It means that during the inelastic cotunneling the state of the quantum dot changes $|i\rangle \neq |f\rangle$. So one can call this process as incoherent~\cite{Footnote}. 
Elastic co-tunneling process can  either change the state of the dot  ($|i\rangle \neq |f\rangle$  the electron spin flips) or does not  $|i\rangle = |f\rangle$.  
One can refer to the latter type of co-tunneling as coherent. In what follows each term in the current will be discussed according to the definitions introduced above.  

\section{Co-tunneling current \label{Sec_ccurrent}}
\subsection{General expression}

Working out Eq.~\eqref{Current4} we obtain the following  expression for the current
\begin{gather}
I_{nin}^{(4)}=\frac{2}{Z_2}\int\frac{d\varepsilon_1d\varepsilon_2}{(2\pi)^2}\Bigl [
(1-n^l_F(\varepsilon_1))n^r_F(\varepsilon_2)\chi^{<,>}(\varepsilon_1,\varepsilon_2)\notag +\\+(1-n^l_F(\varepsilon_1))(1-n^r_F(\varepsilon_2))\chi^{<,<}(\varepsilon_1,\varepsilon_2) + \notag \\
+n^l_F(\varepsilon_1)(1-n^r_F(\varepsilon_2))\chi^{>,<}(\varepsilon_1,\varepsilon_2)+\notag \\+n^l_F(\varepsilon_1)n^r_F(\varepsilon_2)\chi^{>,>}(\varepsilon_1,\varepsilon_2) \Bigr ].
\label{Inin}
\end{gather}
Here $n_F^{l,r}(\varepsilon)$ stands for the Fermi-Dirac distribution of electrons in left/right lead.
Terms in Eq.~\eqref{Inin} proportional to  $(1-n^l_F(\varepsilon_1))(1-n^r_F(\varepsilon_2))$ and $n^l_F(\varepsilon_1)n^r_F(\varepsilon_2)$ have no physical meaning. Therefore  $\chi^{>,>}$  and $\chi^{<,<}$ have to vanish; it is in agreement with direct calculations.
The detailed expressions for $\chi^{<,>}$ and $\chi^{>,<}$ are cumbersome. 
As an example, we present expressions for parts of $\chi^{<,>}$ and $\chi^{>,<}$ which contribute to the non-interference part of the co-tunneling current are presented in Appendix~\ref{App2} (cf. Eqs.~\eqref{ch1}-\eqref{ch4}).
Next,
 \bb
Z_2 = e^{-\beta(\epsilon_1+\epsilon_2)}(e^{\beta \Delta}+e^{-\beta \Delta}+3e^{2\beta J}+1)
\ee
is the canonical partition function for the Hamiltonian $H_{QD}$ with $N=2$ electrons. Analytical expressions for non-interference and interference contributions to $\chi$'s are presented below for antiparallel and parallel alignment of magnetization in the leads. 

\subsection{Antiparallel configuration}

Antiparallel configuration of magnetization in the leads corresponds to the following values of  $g^{l,r}$:
\begin{gather}
\begin{split}
g^l_{1,\uparrow}=g^l_{2,\uparrow}=g^l, \qquad g^r_{1,\downarrow}=g^r_{2,\downarrow}=g^r, \\
g^l_{1,\downarrow}=g^l_{2,\downarrow}=g^r_{1,\uparrow}=g^r_{2,\uparrow}=0.
\end{split}
\end{gather}
This choice of the values of  $g^{l,r}$ assumes that  spin-down electron band of left lead and spin-up electron band of right lead are empty.
In addition we introduce phases $\phi_l$ and $\phi_r$: 
\bb
\overline{t}^l_{1\uparrow}t^l_{2\uparrow}=|\overline{t}^l_{1\uparrow}||t^l_{2\uparrow}|e^{i\phi_l}, \quad
\overline{t}^r_{1\downarrow}t^r_{2\downarrow}=|\overline{t}^r_{1\downarrow}||t^r_{2\downarrow}|e^{i\phi_r} .
\ee

In the anti-parallel case we obtain ($\phi=\phi_r-\phi_l)$
\begin{align}
\chi^{<,>}_{AP}(\varepsilon_1,\varepsilon_2) =& -\frac{4\pi\Delta^2 g^lg^r}{E_c^2}e^{-\beta(\epsilon_1+\epsilon_2-2J)}\Bigl [\notag\\& 2\delta(\varepsilon_1-\varepsilon_2)(1+\cos\phi)
\notag\\+&
\delta(\varepsilon_1-\varepsilon_2+2J)(1-\cos\phi)
\notag \\+&\delta(\varepsilon_1-\varepsilon_2-\Delta+2J)
\notag\\+&
\delta(\varepsilon_1-\varepsilon_2+\Delta+2J)\notag \\+&e^{-2\beta J}\delta(\varepsilon_1-\varepsilon_2-2J)(1-\cos\phi)\notag \\
+&e^{\beta(\Delta-2J)}\delta(\varepsilon_1-\varepsilon_2+\Delta-2J)
\notag\\+&
e^{-\beta(\Delta+2J)}\delta(\varepsilon_1-\varepsilon_2-\Delta-2J)\Bigr ] , \label{chiAP}
\end{align}
and
\begin{equation}
\chi^{>,<}_{AP}(\varepsilon_1,\varepsilon_2) = -  \chi^{<,>}_{AP}(\varepsilon_2,\varepsilon_1) .
\end{equation}
\begin{figure}
\centerline{\includegraphics[height=8cm]{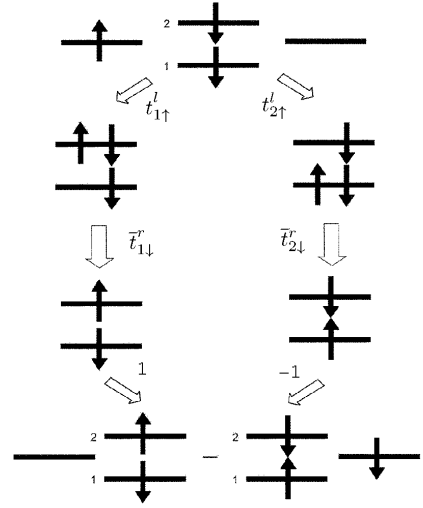}}
\caption{Example of inelastic co-tunneling process that gives contribution to the interference part of the current for anti-parallel configuration. See text.}
\label{Trans}
\end{figure}
Each term in $\chi_{AP}^{<,>}$  
has  transparent physical interpretation; it can be written as
$\sim\exp(-E_i/T) \delta(\varepsilon_1-\varepsilon_2+E_f-E_i)$. For example, the term proportional to $\delta(\varepsilon_1-\varepsilon_2)$ corresponds to the elastic cotunneling of an electron which results in the transition of the quantum dot between the state $|2110\rangle$ and  $|211-1\rangle$. 
Similarly, the term in Eq.~\eqref{chiAP} proportional to $\delta(\varepsilon_1-\varepsilon_2-\Delta+2J)$ describes inelastic cotunneling and corresponds to the transition of the quantum dot from the state $|211-1\rangle$ to $|2020\rangle$. This transition can be realized through two virtual states: with three electrons (upper path) and one electron (lower path) on the dot. Under our assumption of large  $E_c$ both of them provide equal contribution to the current: the energies of the virtual states are  equal to  $E_c$ with our accuracy. Non-zero value of exchange energy $J$ allows the terms in $\chi^{<,>}$ (proportional to $\delta(\varepsilon_1-\varepsilon_2\pm 2J)$) which are inelastic but depends on phase difference $\phi$. They correspond to transitions between the states $|2110\rangle$ and $|211\pm1\rangle$ (see Fig.~\ref{Trans}).

Utilizing  Eq.~\eqref{Inin} we find 
\begin{align}
I^{(4)}_{AP}&=-\frac{2\Delta^2 g^lg^r}{\pi Z_2 E_c^2}e^{-\beta(\epsilon_1+\epsilon_2-2J)}\Bigl [2F(V)(1+\cos\phi) \notag \\
&+
 F(V+2J)(1-\cos \phi)
+F(V-\Delta+2J) \notag 
\\&+F(V+\Delta+2J)\notag
+e^{-2\beta J}F(V-2J)(1-\cos\phi) 
\\&+e^{\beta(\Delta-2J)}F(V+\Delta-2J) \notag
\\&+ e^{-\beta(\Delta+2J)}F(V-\Delta-2J)-(V\rightarrow -V)\Bigr ]\notag \\
&= I_{AP,nin}^{(inel)}+ I_{AP,nin}^{(el)}+ I_{AP,in}^{(inel)}+ I_{AP,in}^{(el)}
\label{cuinap}
\end{align}
where $F(\varepsilon)=\varepsilon /[\exp(\varepsilon/T)-1]$. The inelastic and elastic contributions of non-interference part of the co-tunneling current are given as 
\begin{align}
I_{AP,nin}^{(inel)}&= - \frac{2\Delta^2 g^lg^r}{\pi Z_2 E_c^2}e^{-\beta(\epsilon_1+\epsilon_2-2J)}\Bigl [F(V+2J)\notag\\
&+ F(V-\Delta+2J)+F(V+\Delta+2J)\notag\\
&+ e^{-2\beta J}F(V-2J)+e^{\beta(\Delta-2J)}F(V+\Delta-2J)\notag\\
&+ e^{\beta(-\Delta-2J)}F(V-\Delta-2J)-(V\rightarrow -V)\Bigr ]   , \label{inel/nin}
\end{align}
and
\begin{eqnarray}
I_{AP,nin}^{(el)}=-\frac{4g^lg^r  \Delta^2}{\pi  E_c^2 Z_2} e^{-\beta(\epsilon_1+\epsilon_2-2J)}V , \label{el/nin}
\end{eqnarray}
respectively. Here we use the following relation: $F(V)-F(-V)=-V$. The interference term of the co-tunneling current is splitted on inelastic and elastic parts as follows
\begin{eqnarray}
I_{AP,in}^{(inel)}&=&\frac{2\Delta^2 g^lg^r}{\pi Z_2 E_c^2}e^{-\beta(\epsilon_1+\epsilon_2-2J)}\Bigl [F(V+2J)\cos \phi\notag\\
&+&e^{-2\beta J}F(V-2J)\cos\phi-(V\rightarrow -V) \Bigr ] 
\label{inel/in}
\end{eqnarray}
and
\begin{eqnarray}
I_{AP,in}^{(el)}=- \frac{4g^lg^r  \Delta^2}{\pi  Z_2 E_c^2} e^{-\beta(\epsilon_1+\epsilon_2-2J)}V \cos\phi. \label{el/in}
\end{eqnarray}

As we have mentioned above some inelastic transitions (e.g. from  $|2010\rangle$ to $|211-1\rangle$) can be implemented through two different virtual states. The interference of these two processes depends on the phases of tunneling amplitudes. Therefore there is the interference term in the current  which involves inelastic contibutions of the form $F(\pm2J+ V)$ (see Eq.~\eqref{inel/in}). In the regime of low temperatures and voltages, $|V|,T,|\Delta-2J|\ll\Delta,J$ the inelastic terms of the interference contribution \eqref{inel/in} are suppresed due to small exponential factor $\exp(-2J/T)$. 
In the case of $J=0$ the contribution $I_{AP,in}^{(inel)}$ becomes elastic and exactly compensate  $I_{AP,in}^{(el)}$. Therefore, at $J=0$ the co-tunneling current becomes independent of the phase $\phi$.  

\begin{figure*}[t]
\centerline{\includegraphics[width=55mm]{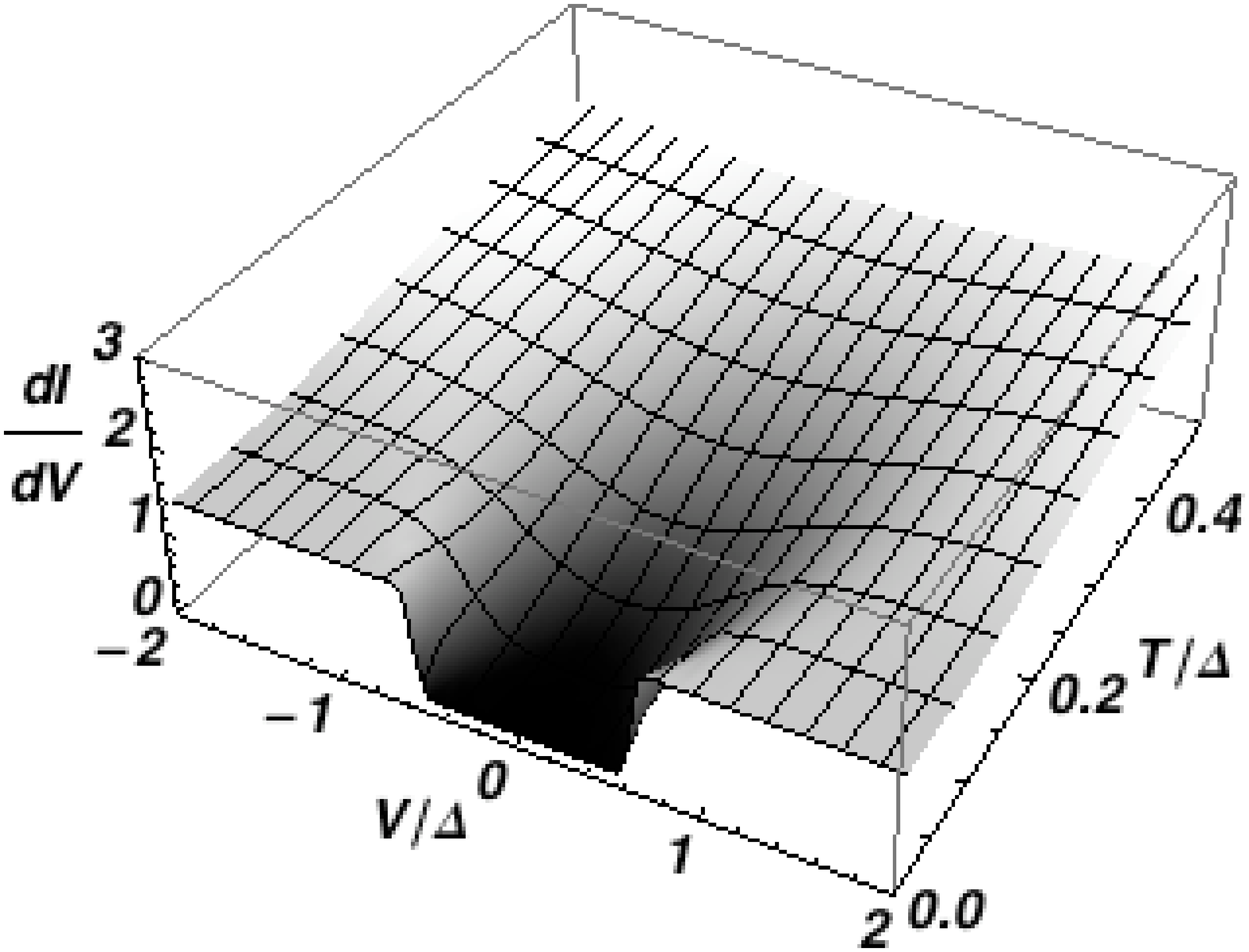}\hspace{1cm} 
\includegraphics[width=55mm]{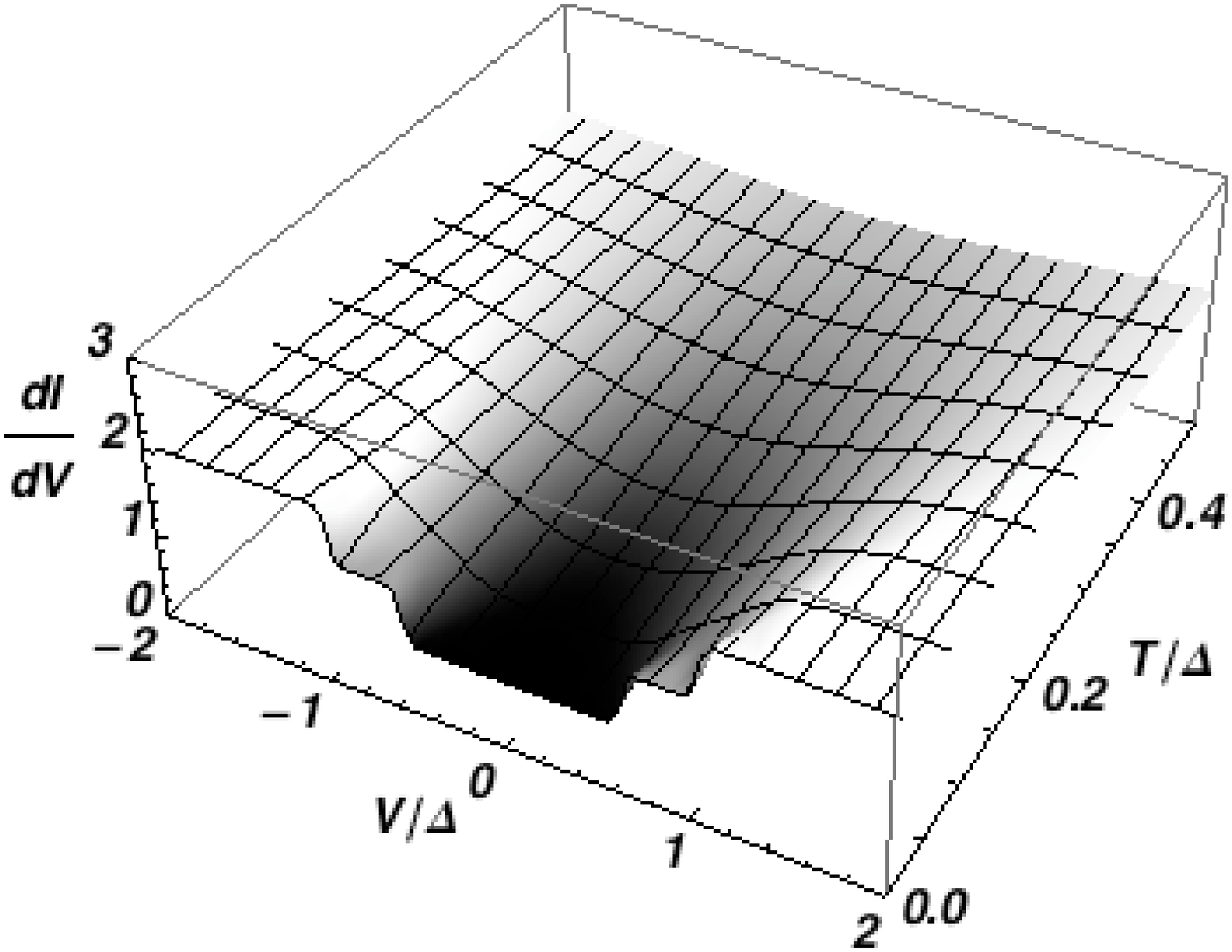}}
\centerline{\includegraphics[width=55mm]{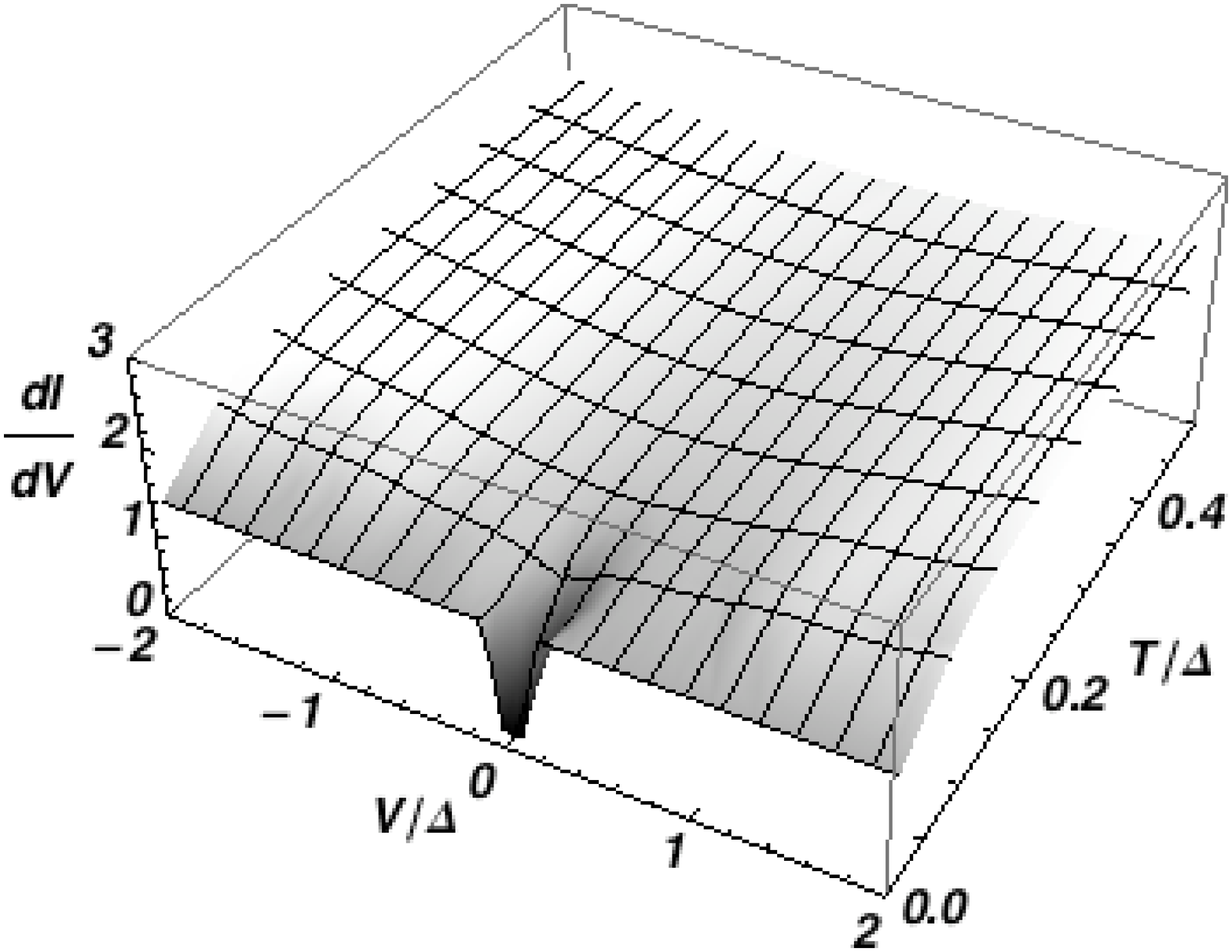}\hspace{1cm}  
\includegraphics[width=55mm]{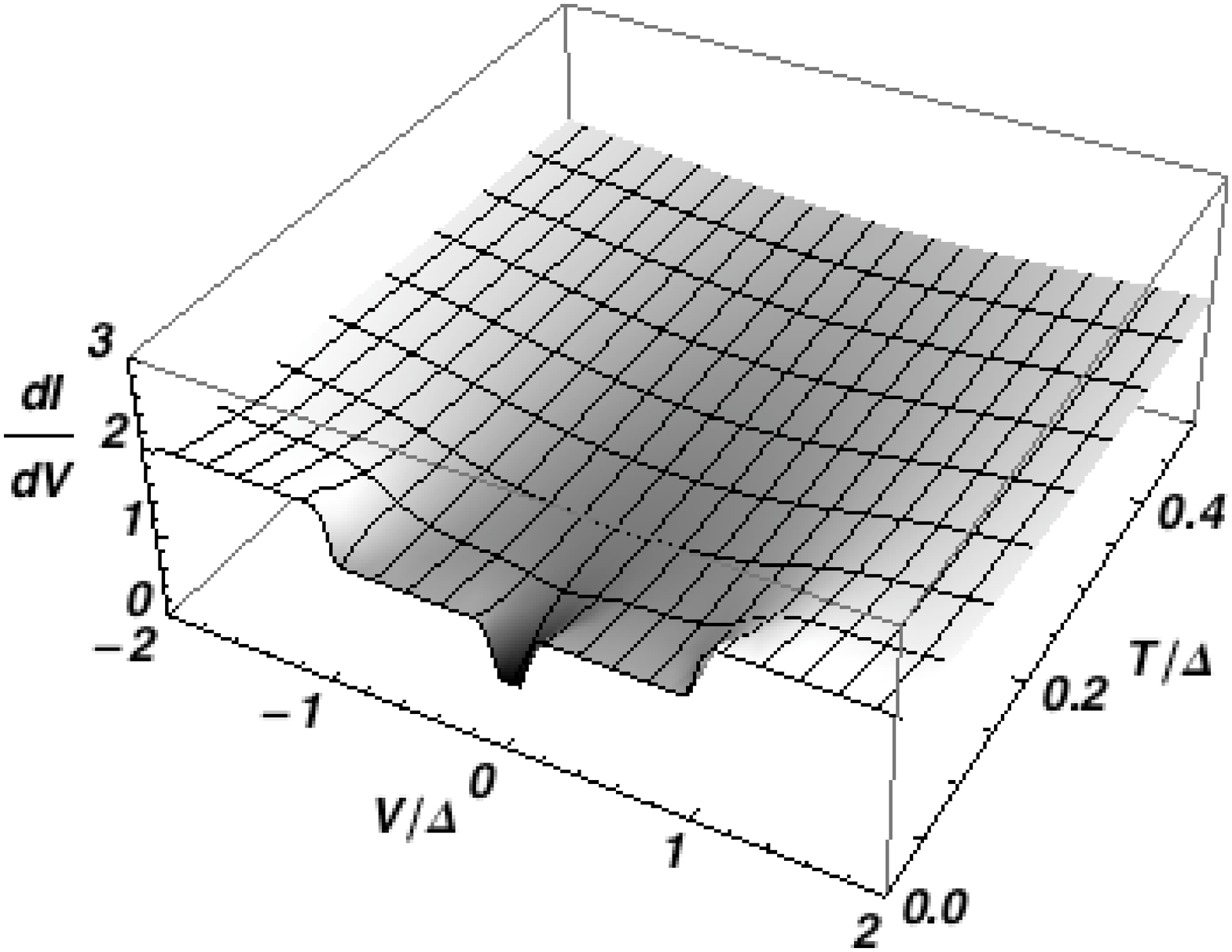}}
\centerline{\includegraphics[width=55mm]{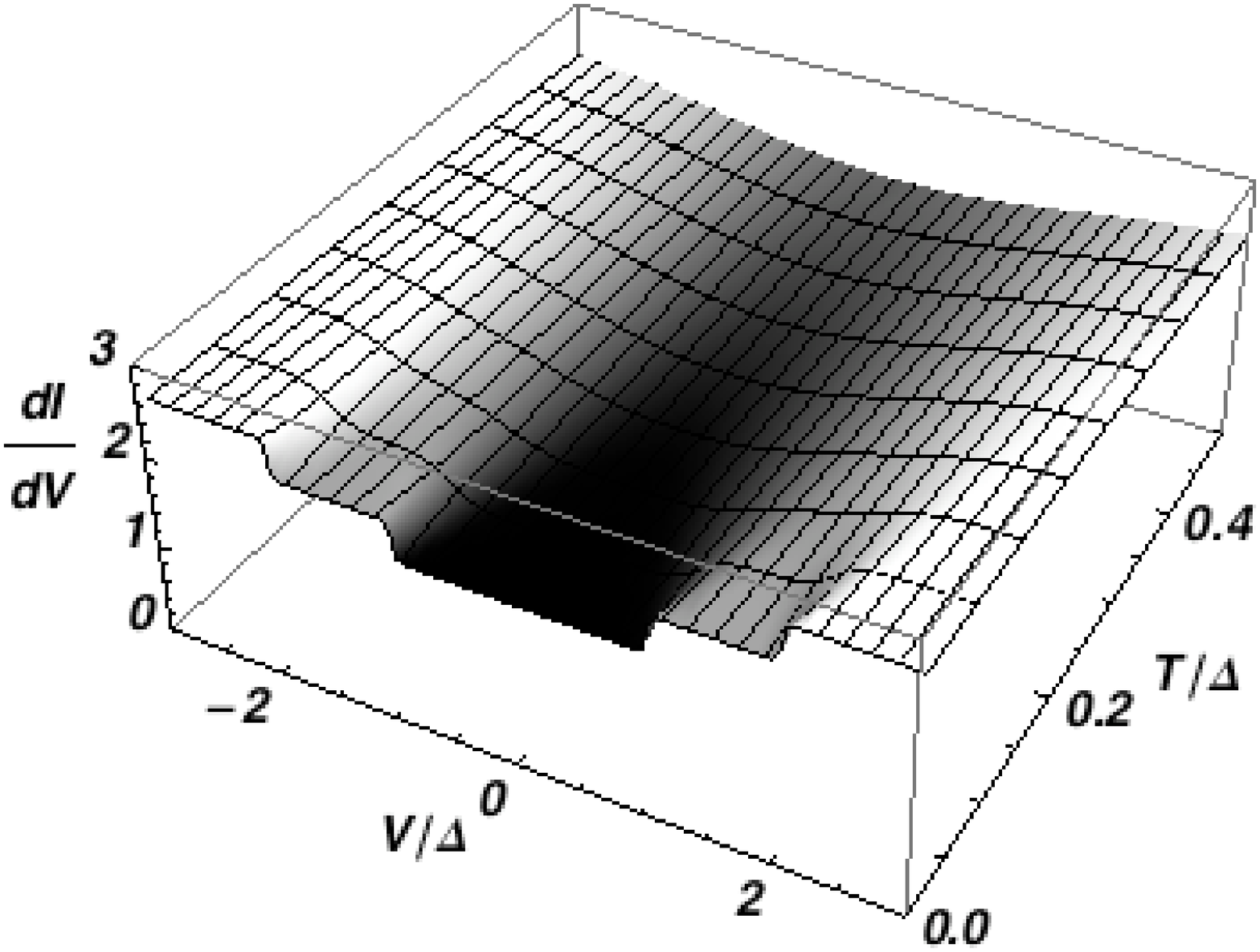}\hspace{1cm}  
\includegraphics[width=55mm]{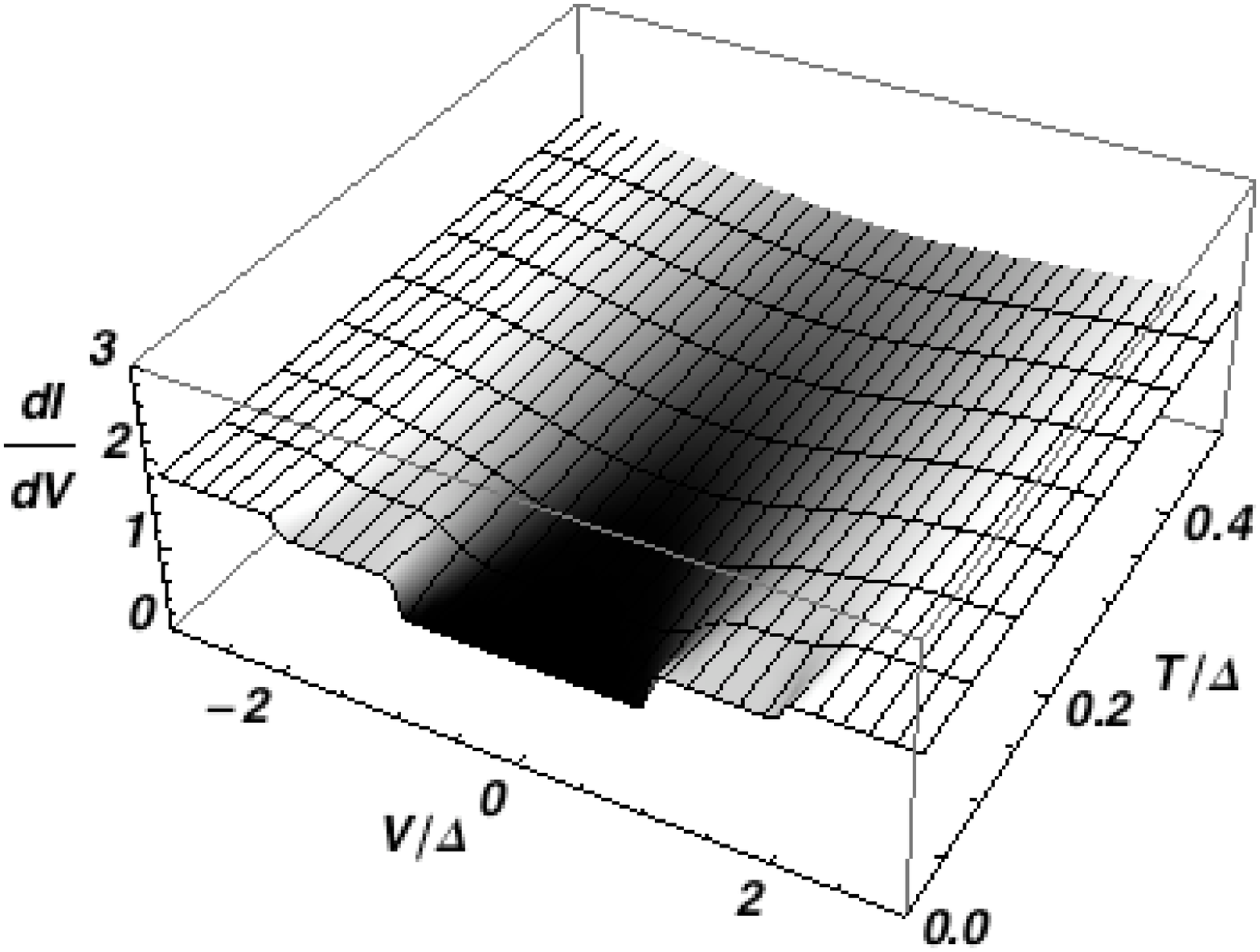}}
\centerline{\includegraphics[width=55mm]{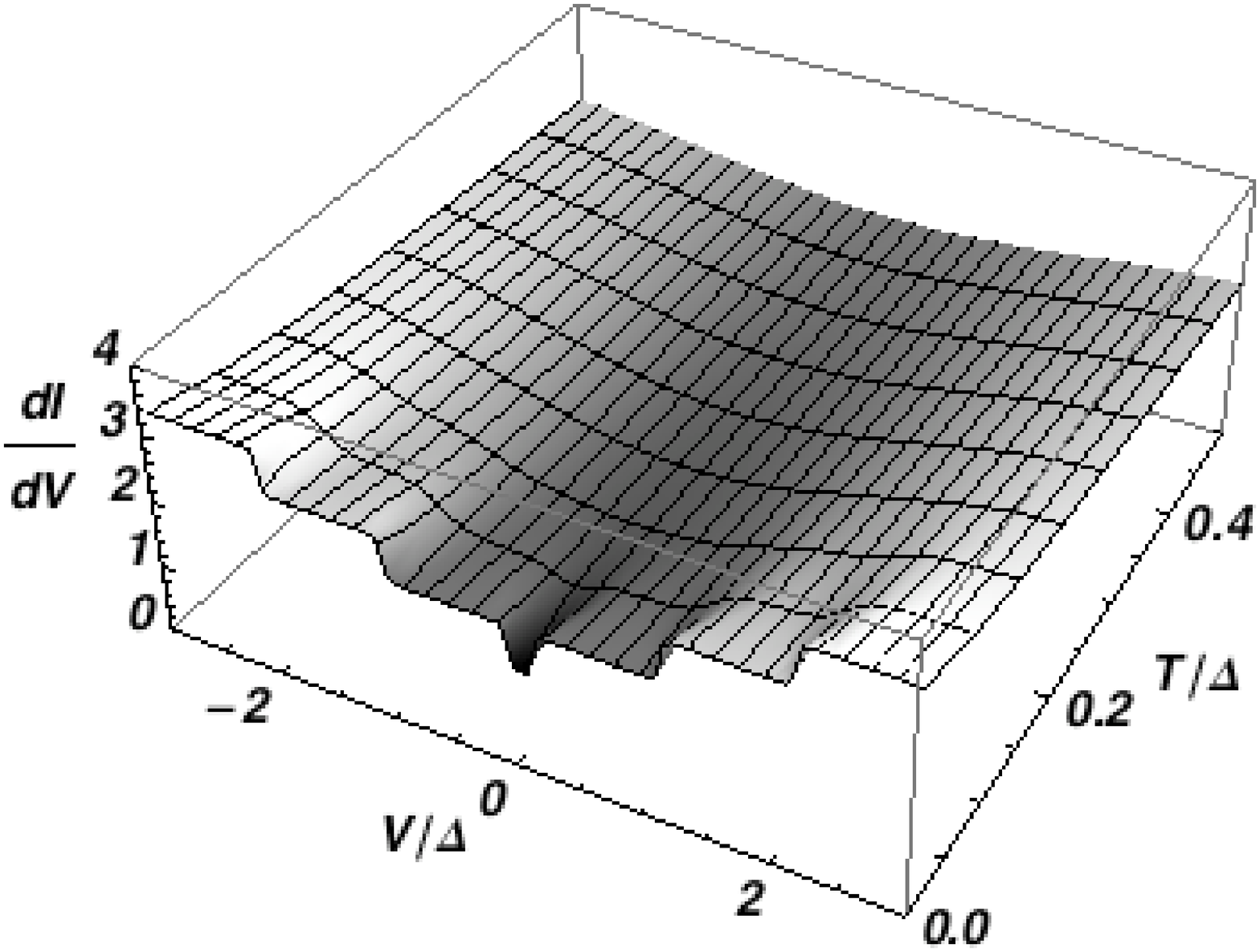}\hspace{1cm}  
\includegraphics[width=55mm]{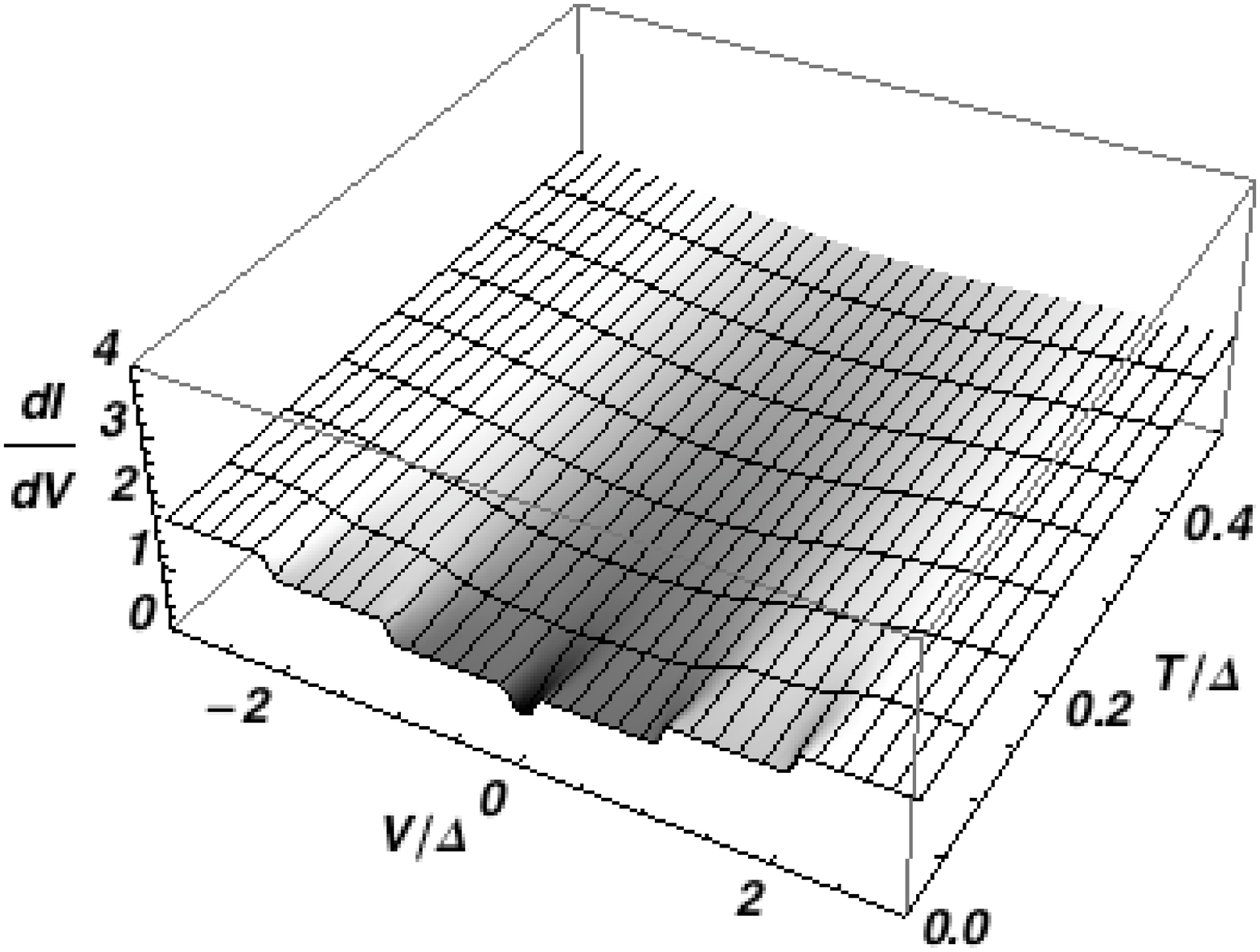}}
\centerline{\includegraphics[width=55mm]{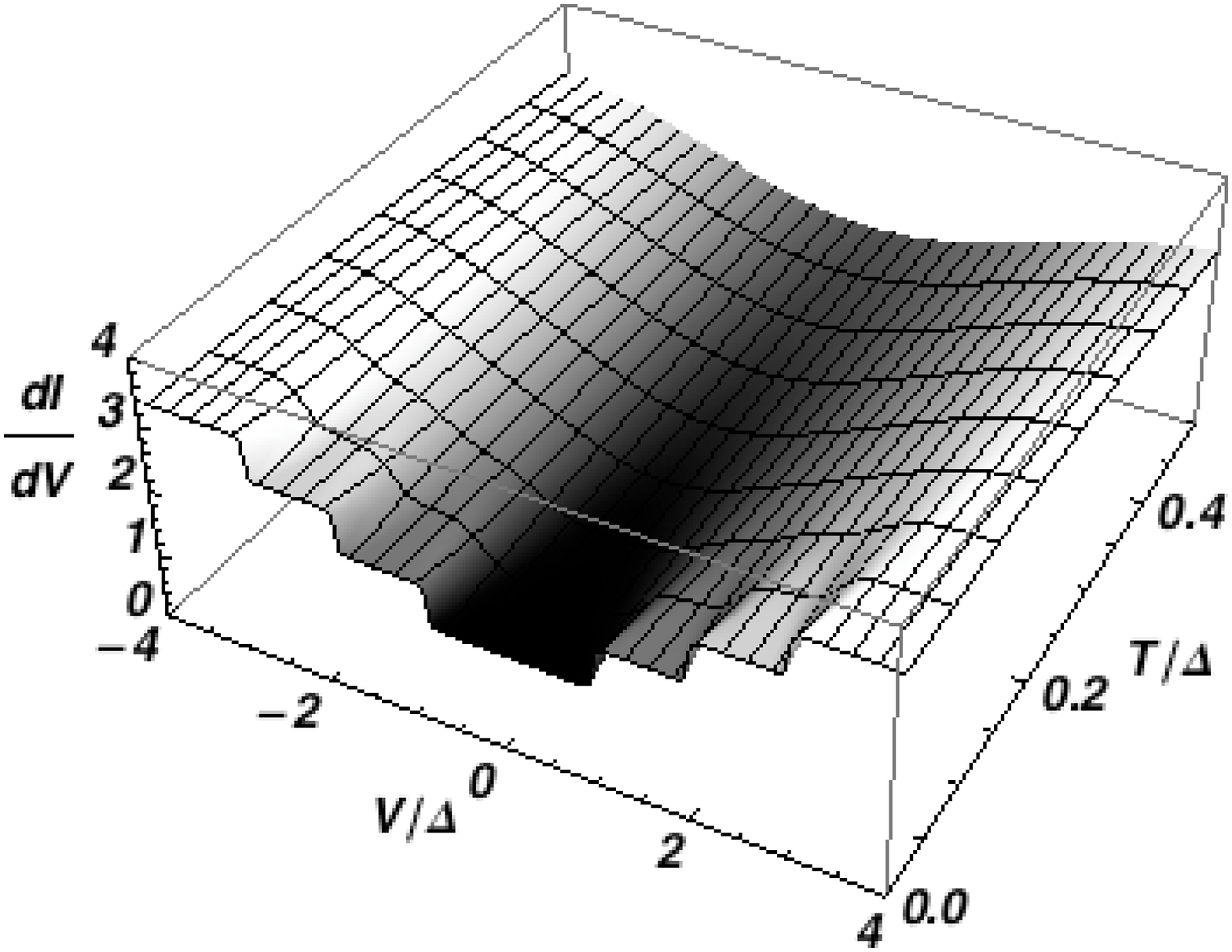}\hspace{1cm}  
\includegraphics[width=55mm]{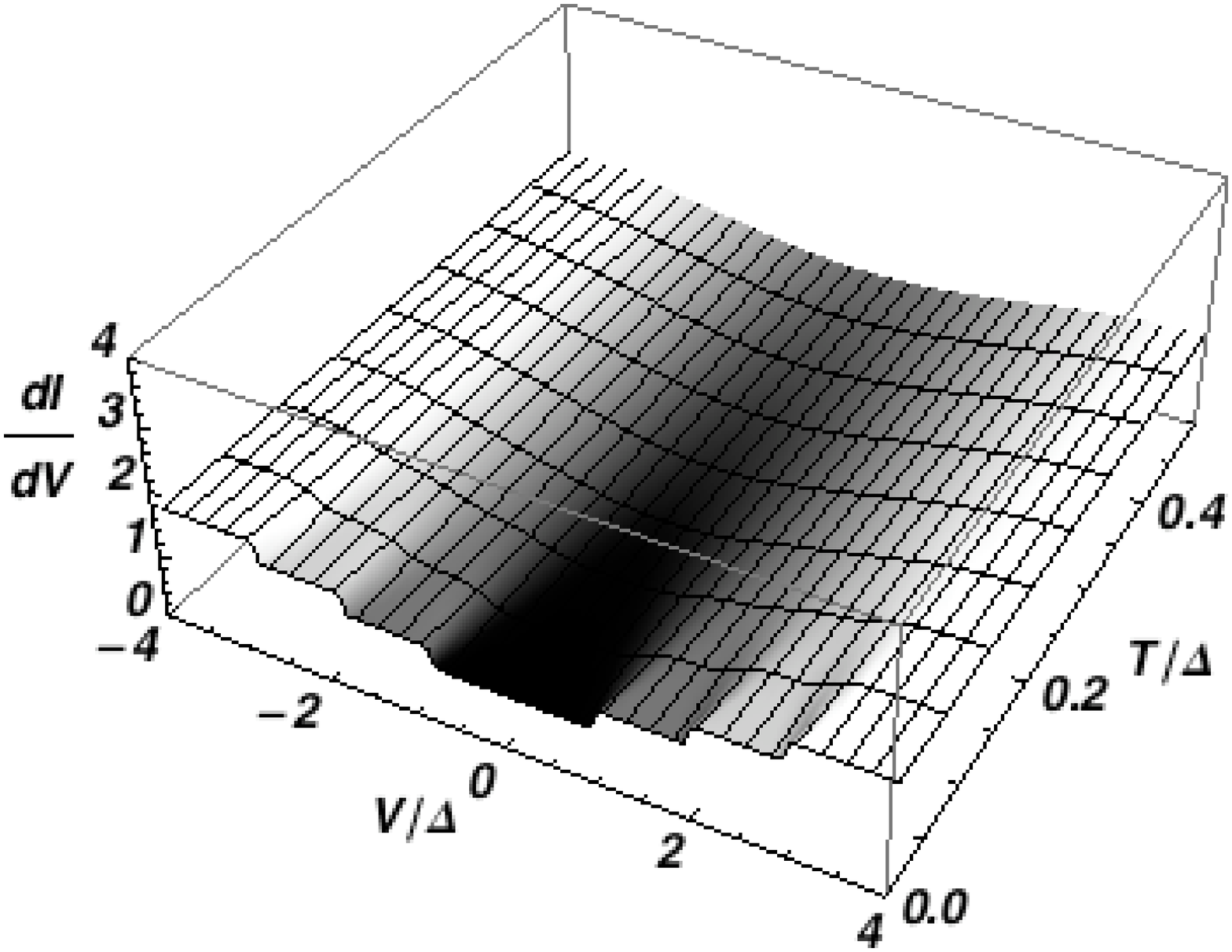}}
\caption{Dependence of the differential conductance $dI^{(4)}/dV$ on $V$ and $T$ for the antiparallel (left) and parallel (right) configurations for different values of $J$. We use $g_l =0.1$, $g_r=0.2$, $E_c/\Delta=10$, $\phi=\pi/2$ and  $J/\Delta=0.2, 0.45, 0.5, 0.55, 0.95$ from top to bottom.}
\label{GraphAP}
\end{figure*}

\begin{figure}
\includegraphics[width=75mm]{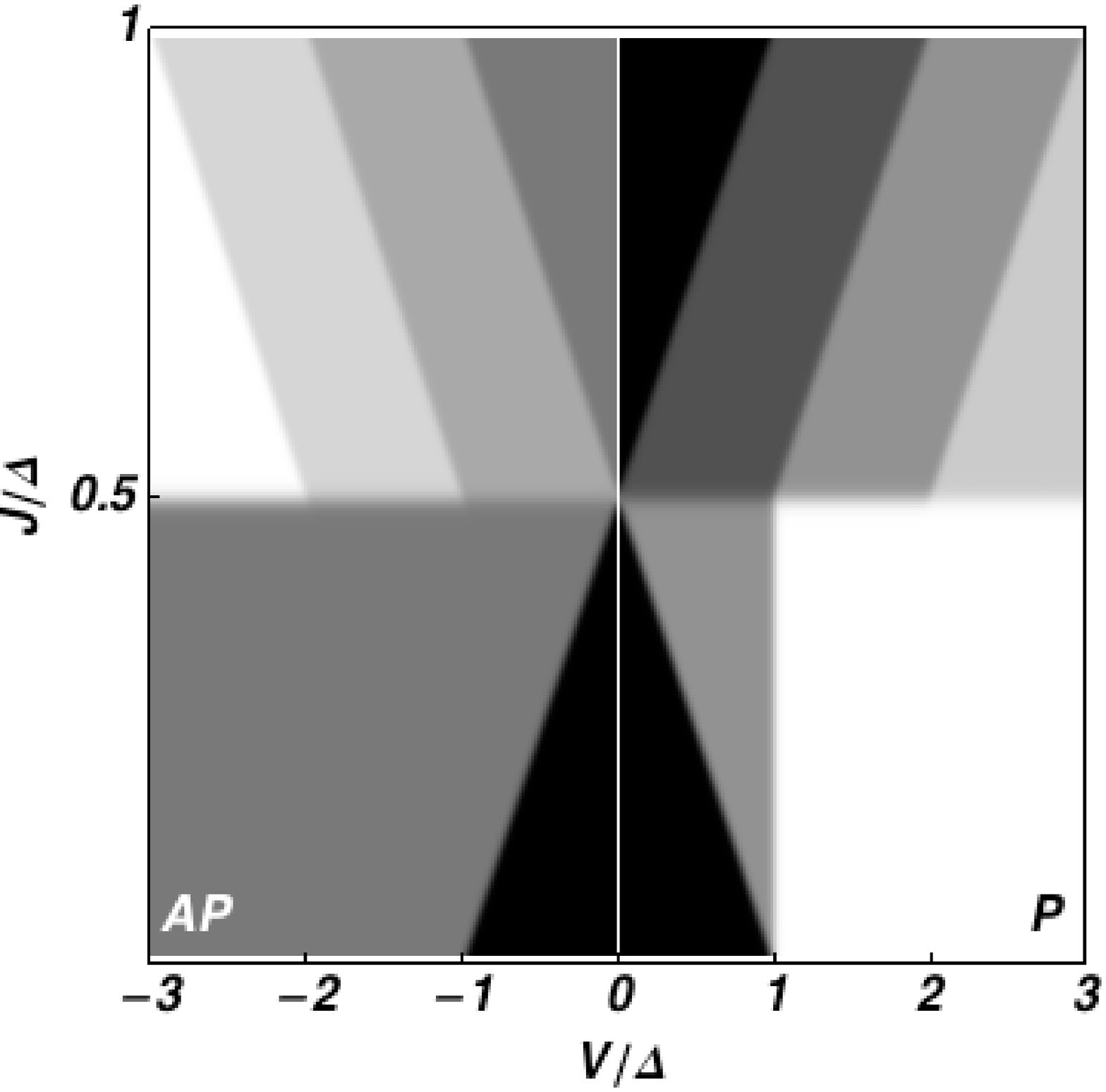}
\caption{Dependence of the differential conductance $dI^{(4)}/dV$ on $V$ and $J$ for the antiparallel (left panel) and parallel (right panel) configurations at $\phi=\pi/2$. We use $g_l =0.1$, $g_r=0.2$, $E_c/\Delta=10$ and $T/\Delta=0.01$.}
\label{Figure_DConJ}
\end{figure}

In Fig.~\ref{GraphAP} we present the dependence of the differential conductance  $dI^{(4)}_{AP}/dV$ on $T$ and $V$ for $\phi=\pi/2$ and different values of the exchange interaction $J$. At low temperatures, $T\ll J, \Delta$ the stair-case structure in differential conductance appears. It corresponds to the step-by-step switching on of the different inelastic processes with increasing voltage. As follows from Eq.~\eqref{cuinap}, one can expect features at $|V|=|\Delta-2J|, 2J, \Delta+2J$. However some of steps have exponentially small height at low temperatures and, therefore, are invisible as shown in Fig.~\ref{GraphAP}. At $J=0$ there is only one stair-case which corresponds to change of the energy of the quantum dot by $\Delta E_{QD}=\Delta$.  For small values of the exchange energy, $J\lesssim \Delta/2$, the feature at $|V|=\Delta E_{QD}=\Delta-2J$  is visible. In the regime $\Delta-2J\ll \Delta, J$, the stair-cases corresponding to processes with $\Delta E_{QD}= \Delta+2J$, $\Delta E_{QD}= 2J$ and $\Delta E_{QD}=\Delta-2J$ appear. 
As expected, the latter disappears at $J=\Delta/2$. All three stair-cases survive at $J>\Delta/2$. The evolution of the differential conductance $dI^{(4)}_{AP}/dV$ with increasing $J$ at fixed temperature is shown in Fig.~\ref{Figure_DConJ} for $\phi=\pi/2$. At $J>\Delta/2$ the feature corresponding to the inelastic process with $\Delta E_{QD}=2J$ disappears at $\phi=0$.

In the most interesting regime of vicinity of the singlet-triplet transition where $\Delta=2J-\kappa$ with $|V|, T, |\kappa|\ll\Delta, J$ the expression for $I_{AP,nin}^{(inel)}$ can be drastically simplified:  
\begin{eqnarray}
I_{AP,nin}^{(inel)}&=&\frac{2g^lg^r\Delta^2}{\pi(3+e^{-\kappa/T}) E_c^2}\Bigl[(V+\kappa)\frac{1-e^{V/T}}{1-e^{(V+\kappa)/T}}\notag\\&+&(V-\kappa)\frac{1-e^{-V/T}}{1-e^{(-V+\kappa)/T}}\Bigr] .\label{asapeq}
\end{eqnarray}
At  $\kappa=0$  the current \eqref{asapeq} acquires particularly simple form:  $I_{AP,nin}^{(inel)}= g^lg^r \Delta^2 V/(\pi E_c^2)$, since the spin flip process becomes elastic. 
  This is the reason why the conductance does not turn into zero when $V=0$.
For  $|\kappa|\gg T$, the curent $I_{AP,nin}^{(inel)}$ becomes 
\begin{eqnarray}  
         I_{AP,nin}^{(inel)}&=&\frac{2c_{\kappa}\Delta^2g^lg^r}{\pi E_c^2}  e^{-|\kappa|/T }
         \notag \\
         &\times&
           \begin{cases}  
   2(|\kappa| \sh \frac{V}{T}+V[1-\ch\frac{V}{T}]),&\quad|V|<|\kappa| ,
 \\ 
 e^{|\kappa|/T }(V-|\kappa|\sgn(V)), & \quad
 |V|>|\kappa| ,
           \end{cases} \notag \\
               \label{asap}
\end{eqnarray}
where  
\bb
c_\kappa=\begin{cases}
  1/3, & \quad \kappa>0 ,\\
   1,& \quad \kappa<0 .
  \end{cases}
\ee
Expression~\eqref{asap} demonstrates exponential suppression of conductance at low temperatures due to the spacing between energy levels (in our case it is spacing between triplet and singlet energy levels which is equal to $|\Delta-2J|$).

\subsection{Parallel configuration}

Parallel configuration of the magnetization in the leads corresponds to the following values of $g^{l,r}$:
\begin{gather}
\begin{split}
g^l_{1,\uparrow}=g^l_{2,\uparrow}=g^l, \qquad g^r_{1,\uparrow}=g^r_{2,\uparrow}=g^r ,\\
g^l_{1,\downarrow}=g^l_{2,\downarrow}=g^r_{1,\downarrow}=g^r_{2,\downarrow}=0.
\end{split}
\end{gather}
This choice of the values of  $g^{l,r}$ assumes that  spin-down electron band of both leads are empty.
In addition we introduce phases $\phi_l$ and $\phi_r$: 
\bb
\overline{t}^l_{1\uparrow}t^l_{2\uparrow}=|\overline{t}^l_{1\uparrow}||t^l_{2\uparrow}|e^{i\phi_l}, \quad
\overline{t}^r_{1\uparrow}t^r_{2\uparrow}=|\overline{t}^r_{1\uparrow}||t^r_{2\uparrow}|e^{i\phi_r} .
\ee
In this case we obtain
\begin{align}
\chi_P^{<,>}(\varepsilon_1,\varepsilon_2)=&-\frac{2\pi\Delta^2 g^lg^r}{E_c^2}e^{-\beta(\epsilon_1+\epsilon_2-2J)}\Bigl \{\notag\\&2\delta(\varepsilon_1-\varepsilon_2)(1+\cos\phi)\notag \\+&\notag\delta(\varepsilon_1-\varepsilon_2-2J)(1-\cos\phi)\\+&\notag\delta(\varepsilon_1-\varepsilon_2-\Delta-2J)
+\delta(\varepsilon_1-\varepsilon_2+\Delta-2J)\notag
\\+& e^{-2\beta J}\Bigl [ \delta(\varepsilon_1-\varepsilon_2+2J)(1-\cos\phi)\notag \\+&\notag\delta(\varepsilon_1-\varepsilon_2-\Delta)+\delta(\varepsilon_1-\varepsilon_2+\Delta)\Bigr ]
\notag\\+&e^{\beta(\Delta-2J)}\Bigl [ \delta(\varepsilon_1-\varepsilon_2)(1-\cos\phi)\notag\\+& \delta(\varepsilon_1-\varepsilon_2-\Delta+2J)+\notag\delta(\varepsilon_1-\varepsilon_2-\Delta) \Bigr ]\notag\\+&
e^{-\beta(\Delta+2J)}\Bigl [ \delta(\varepsilon_1-\varepsilon_2)(1-\cos\phi)\notag\\+&\delta(\varepsilon_1-\varepsilon_2+\Delta+2J)+\delta(\varepsilon_1-\varepsilon_2+\Delta)\Bigr ]\Bigr \} ,\label{cdf1}
\end{align}
and 
\begin{equation}
\chi_P^{>,<}(\varepsilon_1,\varepsilon_2)=-\chi_P^{<,>}(\varepsilon_2,\varepsilon_2)
\end{equation}

As in the case of the antiparallel alignment of magnetizations, each term in $\chi_P^{<,>}$ 
has  transparent physical interpretation. As compared with $\chi_{AP}^{<,>}$, Eq.~\eqref{cdf1} demonstrates that the case of parallel  magnetizations allows more elastic processes. In addition, transitions in which energy of the quantum dot  is changed by $\Delta E_{QD}=\pm \Delta$ are possible.

Utilizing Eq.~\eqref{Inin}, we find 
\begin{eqnarray}
I_{P}^{(4)} &=&-\frac{\Delta^2g^lg^r}{\pi Z_2 E_c^2}e^{-\beta(\epsilon_1+\epsilon_2-2J)}\Bigl \{2F(V)(1+\cos\phi)\notag\\&+& F(V+2J)(1-\cos\phi)+F(V-\Delta+2J)\notag\\&+&F(V+\Delta+2J)\notag\\&+&e^{-2\beta J}\Bigl [ F(V-2J)(1-\cos\phi)\notag\\&+&F(V+\Delta)+F(V-\Delta)\Bigr ]\notag\\
&+&e^{\beta(\Delta-2J)}\Bigl [ F(V)(1-\cos\phi)\notag\\&+&F(V+\Delta-2J)+F(V+\Delta)\Bigr ]\notag\\&+&
e^{-\beta(\Delta+2J)}\Bigl [F(V)(1-\cos\phi)\notag \\&+&F(V-\Delta-2J)\notag\\&+&F(V-\Delta)\Bigr ]-
(V\rightarrow-V)\Bigr\}\notag\\
&=& I_{P,nin}^{(inel)}+ I_{P,nin}^{(el)}+ I_{P,in}^{(inel)}+ I_{P,in}^{(el)} . 
\label{CurrentP}
\end{eqnarray}
The inelastic and elastic parts of the non-interference contribution to the co-tunneling current are as follows:
\begin{eqnarray}
I_{P,nin}^{(inel)}&=&-\frac{\Delta^2g^lg^r}{\pi Z_2 E_c^2}e^{-\beta(\epsilon_1+\epsilon_2-2J)}\Bigl \{  F(V+2J)\notag\\
&+&F(V-\Delta+2J)+F(V+\Delta+2J)\\&+&e^{-2\beta J}\Bigl [F(V-2J)+F(V+\Delta)+F(V-\Delta)\Bigr ]
\notag\\&+&e^{\beta(\Delta-2J)}\Bigl [F(V+\Delta-2J)+F(V+\Delta)\Bigr ]\notag\\&+&
e^{-\beta(\Delta+2J)}\Bigl [+F(V-\Delta-2J)\notag\\&+&F(V-\Delta)\Bigr ]-
(V\rightarrow-V) \Bigr \}  \label{inel/ninP}
\end{eqnarray}
and
\begin{eqnarray}
I_{P,nin}^{(el)}&=&\notag-\frac{2\Delta^2g^lg^r}{\pi Z_2 E_c^2}e^{-\beta(\epsilon_1+\epsilon_2-2J)} \Bigl (2+e^{-\beta(\Delta+2J)}\\&+&e^{\beta(\Delta-2J)}\Bigr ) V  . \label{el/ninP}
\end{eqnarray}
The inelastic and elastic terms for the interference part of the co-tunneling current are given by
\begin{eqnarray}
I_{P,in}^{(inel)}&=&\frac{\Delta^2g^lg^r}{\pi Z_2 E_c^2}e^{-\beta(\epsilon_1+\epsilon_2-2J)}\Bigl \{F(V+2J)\notag\\&+ &e^{-2\beta J}F(V-2J)
\notag\\&-&
(V\rightarrow-V) \Bigr \} \cos \phi
\label{inel/inP}
\end{eqnarray}
and
\begin{eqnarray}
I_{P,in}^{(el)}&=&\notag -\frac{2g^lg^r  \Delta^2}{\pi Z_2 E_c^2}e^{-\beta(\epsilon_1+\epsilon_2-2J)}
\Bigl [2-e^{\beta(\Delta-2J)}\\&-&e^{-\beta(\Delta+2J)} \Bigr ] V \cos\phi  . \label{el/inP}
\end{eqnarray}

In the expression~\eqref{CurrentP} there are two types of additional terms in comparison with the case of the antiparallel configuration.  First ones ($\propto F(V\pm\Delta))$) correspond to the singlet-singlet transitions of the quantum dot (e.g. $|2020\rangle\rightarrow|2010\rangle$) during inelastic co-tunneling  including a transfer of one electron to the different level. Second one ($\propto F(V)(1-\cos\phi)$) are elastic terms due to transitions of the quantum dots between the states in which one-level is doubly occupied, e.g. $|2020\rangle$. These elastic terms lead to dependence of the co-tunneling current for parallel alignment of magnetizations on the phase difference $\phi$ even at $J=0$.

In Fig.~\ref{GraphAP} we present the dependence of the differential conductance  $dI^{(4)}_{P}/dV$ on $T$ and $V$ for $\phi=\pi/2$ and different values of the exchange interaction $J$. At low temperatures, $T\ll J, \Delta$ the stair-case structure in differential conductance appears. It corresponds to the step-by-step switching on of the different inelastic processes with increasing voltage. As follows from Eq.~\eqref{CurrentP}, one can expect features at $|V|=|\Delta-2J|, 2J, \Delta, \Delta+2J$. However some of steps have exponentially small height at low temperatures and, therefore, are invisible as shown in Fig.~\ref{GraphAP}. At $J=0$ there is only one stair-case which corresponds to change of the energy of the quantum dot by $\Delta E_{QD}=\Delta$.  For small values of the exchange energy, $J\lesssim \Delta/2$, additional feature at $|V|=\Delta E_{QD}=\Delta-2J$  is visible. In the regime $\Delta-2J\ll \Delta, J$, the stair-cases corresponding to processes with $\Delta E_{QD}= \Delta+2J$, $\Delta E_{QD}= 2J$ and $\Delta E_{QD}=\Delta-2J$ appear. As expected, the latter disappears at $J=\Delta/2$. As in the case of antiparallel alignment of magnetizations three stair-cases 
at $|V|= 2J-\Delta, 2J, 2J+\Delta$ survive at $J>\Delta/2$. The feature corresponding to $\Delta E_{QD}= \Delta$ becomes invisible at $J>\Delta/2$. The evolution of the differential conductance $dI^{(4)}_{P}/dV$ with increasing $J$ at fixed temperature is shown in Fig.~\ref{Figure_DConJ} for $\phi=\pi/2$. At $J>\Delta/2$ the feature corresponding to the inelastic process with $\Delta E_{QD}=2J$ disappears at $\phi=0$.

In the most interesting regime near the singlet-triplet transition $\Delta=2J-\kappa$: $|V|, T, |\kappa|\ll\Delta, J$ the expression for $I_{P,nin}^{(inel)}$ can be written as
\begin{eqnarray}
I_{P,nin}^{(inel)}&=&-\frac{\Delta^2 g^lg^r}{\pi E_c^2(3+e^{-\kappa/T})}\Bigl[(V-\kappa) \frac{1-e^{-V/T}}{1-e^{(-V+\kappa)/T}}
\notag\\
&+& (V+\kappa) \frac{1-e^{V/T}}{1-e^{(V+\kappa)/T}}
\Bigr] .
\label{aspeq}
\end{eqnarray}
We mention that in this regime $I_{P,nin}^{(inel)} = (1/2)I_{AP,nin}^{(inel)}$. 
In the case  $|\kappa|, |V|, T\ll \Delta, J$  additional terms $\propto F(V\pm\Delta)$ 
are suppresed by the small factor $\exp(-\Delta/T)$ and therefore they do not contribute to the current.

\section{Discussions and conclusions\label{Sec_Conc}}

In this paper the co-tunneling current through the two-level quantum dot coupled to ferromagnetic leads is calculated analytically. The results have been presented for the most interesting case of the quantum dot with two electrons and for parallel and antiparallel configurations of magnetization in the leads.

Inelastic co-tunneling current  has features corresponding to transitions between energy levels of the quantum dot which could be used to determine structure of energy levels on the quantum dot. In both cases of parallel and antiparallel configurations the non-interference part of the conductance has a minimum at low temperatures and voltages $|V|, T\ll \Delta$.
Width of this minimum near the transition $(\Delta=2J)$ between singlet and triplet ground states  is defined by  the gap between singlet and triplet two-electron states $|\Delta-2J|$. 
Near the transition our results for the differential conductance resembles the expression derived in the paper~\cite{KangMin} for the two-level quantum dot at $J=0$, if we substitute the averaged single-particle spacing by the singlet-triplet gap $|\Delta-2J|$. In the case of low temperatures $T\ll|\Delta-2J|$ inelastic part of the current is suppressed by the factor $\exp(-|\Delta-2J|/T)$ in comparison with elastic one. 
However at temperatures  $T\sim |\Delta-2J|\ll \Delta$ elastic and inelastic parts of the co-tunneling current are of the same order of magnitudes. It is due to the fact that the spin flip processes become almost elastic near the singlet-triplet transition in contrast to the case of $J=0$.

As we mentioned in the Introduction numerical calculations of the differential conductance based on a rate equation approach
were performed in Ref. ~\cite{Weymann}. It was found that there is a zero-bias peak with the width of the order of $T$  in the antiparallel case (in contrast to the parallel one). This zero-bias peak has been explained by non-equilibrium difference in the occupation probabilities of states $|2111\rangle$ and $|211-1\rangle$. The absence of such effects in perturbation theory may be a reason for discordance between our findings and results of Ref.~\cite{Weymann} at small bias. However, we emphasize that both perturbation theory and rate equation method produce the same results for conductance and for the positions of ``stairs" in dependence of the differential conductance on voltage.

In the regime $V, T\sim|\Delta-2J|\ll \Delta$ only singlet-triplet transitions are important. It is worthwhile to mention that in this regime the inelastic interference contributions to the co-tunneling current are exponentialy supressed. Thus, in this regime expressions~\eqref{asap} and \eqref{aspeq} for inelastic part of the co-tunneling current $I^{(inel)}$ are valid for a quantum dot with large number of levels. Elastic part of the co-tunneling current $I^{(el)}$ in such quantum dots is determined by transitions through energy levels in the range $\sim E_c$. Therefore, expression for $I^{(el)}$ in quantum dots with large number of levels is $\sim E_c/\Delta$ times greater than $I^{(el)}$ for a two-level quantum dot (see Eqs.~\eqref{el/nin}, \eqref{el/in}, \eqref{el/ninP} and \eqref{el/inP}). It is this way the latter matches with the result for the elastic co-tunneling in a multi-level quantum dot. 

In quantum dots with large number of levels transitions between ground states with $S$ and $S+1$ are possible at $J=J_S = \Delta (2S+1)/(2S+2)$~\cite{Kurland}. Our results indicate that in vicinity of such transitions at $|V|, T\ll |J-J_S|\ll J, \Delta$ the inelastic part of the co-tunneling current will be suppressed. However, exactly at the transition ($J=J_S$) $I^{(inel)}$ become linear in $V$ and independent of $T$. Therefore, at the transition point the increase of conductance should occur. One can utilize this fact to experimentaly observe the transition between ground states with $S$ and $S+1$ in multi-level quantum dots. 


It is worthwhile to emphasize that the interference part of the co-tunneling current ~\eqref{inel/in} and \eqref{inel/inP} involves terms corresponding to the inelastic processes  $(\propto F(eV\pm2J))$.
This is the consequence of the presence of the two-particle eigenstates  $|211-1\rangle$ and $|2110\rangle$  which allow the quantum dot to make inelastic transitions using different paths. As usual it leads to the dependence of the probabilities of such transitions on phases of the tunneling amplitudes.
This fact differs problem under consideration from the standard one~\cite{NazarovAverinOdintsov} in which only elastic terms depend on phases of the tunneling amplitudes. 
Low-temperature current-voltage characteristics are non-linear but they became linear if the gap $|\Delta-2J|$ tends to zero. One can utilize this fact to observe the quantum dot ground state transition point experimentally. 

To realize effectively two-level quantum dot one can use any system with doubly degenerate levels and study it at voltages and temperatures much less than level spacing. For example, it can be a carbon nanotube with orbital degeneracy of levels~\cite{Paaske}  or a two-dimensional electron gas in Si(001)-MOSFET~\cite{SiQD1} and Si/SiGe heterostructures~\cite{SiQD2}. Also one will observe two-level quantum dot behaviour of inelastic co-tunneling current for arbitrary quantum dot in the regime $|V|, T\sim |J-J_S|\ll J, \Delta$.

The authors thank A. Ioselevich for useful discussions. The research was funded in part by the Russian Ministry of Education and Science under Contract No. P926, RFBR Grants No. 09-02-92474-MHKC, the Council for grants
of the Russian President Grant No. MK-296.2011.2, the Dynasty foundation and RAS Programs ``Quantum Physics of Condensed Matter'' and ``Fundamentals
of nanotechnology and nanomaterials''.

\appendix

\section{General form of Hamiltonian for a two-level quantum dot\label{App1}}

Although there is vast body of studies on spectra in a few electron quantum dots~\cite{FE}, in this appendix we present the general expression for Hamiltonian for a two-level quantum dot and discuss under which assumptions it can be simplified to Eq.~\eqref{Hqd}.
We start from the following Hamiltonian
\bb
H_{QD}=\sum_{\alpha,\sigma}\epsilon_\alpha d^\dag_{\alpha\sigma}d_{\alpha\sigma} + H_{\rm int} \label{ae1}
\ee
where $\alpha=1,2$ denotes the orbital single-particle levels and 
\bb
H_{\rm int} =\frac{1}{2}\sum_{\sigma_1,\sigma_2,\alpha_j}U_{\alpha_1\alpha_2\alpha_3\alpha_4}d^{\dag}_{\alpha_1\sigma_1}d^{\dag}_{\alpha_2\sigma_2}d_{\alpha_3\sigma_2}d_{\alpha_4\sigma_1}
\ee
is the interaction part of Hamiltonian. The matrix elements of interaction are defined as: 
\begin{align}
U_{\alpha_1\alpha_2\alpha_3\alpha_4}&=\int d\bm{r} d\bm{r}^\prime \varphi_{\alpha_1}^*(\bm{r})\varphi_{\alpha_2}^*(\bm{r}^\prime)U(\bm{r}-\bm{r}^\prime)\notag \\
&\times \varphi_{\alpha_3}(\bm{r}^\prime)\varphi_{\alpha_4}(\bm{r})
\end{align}
where $U(r) = e^2/(\varepsilon r)$ is the Coulomb potential. Provided the time-reversal invariance is preserved only six matrix elements $U_{\alpha_1\alpha_2\alpha_3\alpha_4}$ are independent: 
\begin{align}
& U_{1111} ,\notag \\
& U_{1112}=U_{1121}=U_{1211}=U_{2111} ,\notag\\
&U_{1122}=U_{2211} =U_{1212}=U_{2121} ,\notag\\
&U_{1221}=U_{2112} ,\notag\\
&U_{1222}=U_{2122}=U_{2212}=U_{2221} ,\notag\\
&U_{2222} .
\end{align}
Here matrix elements $U_{1111}$, $U_{2222}$ and  $U_{1221}$ describe direct Coulomb interaction whereas $U_{1122}$ corresponds to exchange energy. 

Hamiltonian~\eqref{ae1} commutes with the total number of electrons $\hat N$, the total spin square $\hat{\bm S}^2$, and $S_z$.  Therefore, it is convenient to work in the basis of two-particle states $|NSnS_z\rangle$ presented in Fig.~\ref{FigureLevels}. Then $H_{QD}$ can be written as a $16\times 16$ matrix. All the states except those with $N=2$, $S=0$, $S_z=0$ and with $N=3$, $S=1/2$, $S_z=\pm 1/2$ are eigenstates of Hamiltonian~\eqref{ae1}. Their energies are
\begin{align}
E_{0000} &=0 ,\notag\\
E_{1\frac{1}{2}1\frac{1}{2}} &=
E_{1\frac{1}{2}1-\frac{1}{2}}=\epsilon_1,\quad
E_{1\frac{1}{2}0\frac{1}{2}}=
E_{1\frac{1}{2}0-\frac{1}{2}}=\epsilon_2,\notag \\ 
E_{2111}&=E_{211-1}=E_{2110}=\epsilon_1+\epsilon_2+U_{1221}-U_{1122} ,\notag\\
E_{4020}&= 2\epsilon_1+2\epsilon_2+U_{1111}+U_{2222}+4U_{1221}-2U_{1122}.
\end{align} 
The states $|2010\rangle$, $|2020\rangle$, and $|2000\rangle$ are mixed and Hamiltonian~\eqref{ae1} projected onto these states is given as
\begin{equation}\label{ae4}
H_1 = (\epsilon_1+\epsilon_2)\bm{1} + V_1
\end{equation}
where $\bm{1}$ denotes the unit matrix and 
\begin{equation}\label{ae3}
 V_1=\begin{pmatrix}
U_{1221}+U_{1122}&-\sqrt{2} U_{1112}&-\sqrt{2}U_{1222}\\
-\sqrt{2} U_{1112}&-\Delta+U_{1111}&U_{1122} \\
-\sqrt{2}U_{1222}&U_{1122}&\Delta+U_{2222}
\end{pmatrix} .
\end{equation}
The states $|3\frac{1}{2}2\frac{1}{2}\rangle$ and $|3\frac{1}{2}1\frac{1}{2}\rangle$ are also mixed and Hamiltonian~\eqref{ae1} projected onto these states can be written as
\begin{equation}
H_2 = (\epsilon_1+\epsilon_2+2U_{1221}-U_{1122})\bm{1} + V_2 \label{ae2}
\end{equation}
where 
\begin{equation}
V_2=\begin{pmatrix}
\epsilon_1+U_{1111}&-U_{1112}-U_{1222} \\
-U_{1112}-U_{1222}&\epsilon_2+U_{2222}
\end{pmatrix} .
\end{equation}
Hamiltonian~\eqref{ae2} describes also mixing of the states  $|3\frac{1}{2}2-\frac{1}{2}\rangle$ and $|3\frac{1}{2}1-\frac{1}{2}\rangle$.

As an example, let us consider a quantum dot fabricated in two-dimensional electron gas in Si(001)-MOSFET structure~\cite{SiQD1}. In such quantum dots electrons can occupy states in two valleys which remain from six-fold degeneracy of bulk Si. Assuming the level spacing due to spatial confinement to be large as compared with the valley splitting we have only two  low-energy orbital states (symmetric and anti-symmetric):
\begin{align}
\varphi_1(\bm{r}) &= \sqrt{2} \cos \frac{Q z}{2} \varphi_0(z) \varphi_\perp(\bm{\rho}) ,\notag \\
\varphi_2(\bm{r}) &= \sqrt{2} \sin \frac{Q z}{2} \varphi_0(z) \varphi_\perp(\bm{\rho})  
\end{align}
where $z$ denotes the coordinate perpendicular to the plane in which two-dimensional electron gas is formed, $\bm{\rho}$
 the in-plane coordinate vector, and $\bm{r}=\bm{\rho}+z \bm{e_z}$. The vector $\bm{Q}=(0,0,Q)$ corresponds to the shortest distance between the valley minima in the reciprocal space: $Q\propto 1/a_{\rm lat}$, with $a_{\rm lat}$ being the lattice
constant. We choose the envelope function $\phi_0(z)$ to be normalized and symmetric. We assume that the ground state eigenfunction $\varphi_\perp(\bm{\rho})$ of the confinement potential which creates a quantum dot is also normalized and symmetric. Then, as one can check the matrix elements $U_{1112}=U_{1222}\equiv 0$. Also we assume that the following conditions 
 \begin{equation}
 Q^{-1}\ll l_z \ll l_\perp
 \end{equation}
are satisfied. Here $l_z$ and $l_\perp$ are typical lengths associated with the functions $\varphi_0(z)$ and $\varphi_\perp(\bm{\rho})$: $l^{-1}_z\sim \int dz \,\varphi_0^4(z)$, $l^{-2}_\perp \sim \int d\bm{\rho}\, \varphi_\perp^4(\bm{\rho})$. Then neglecting exponentially small in $Q l_z$ contributions of a type $\int dz\, \varphi_0^4(z) \cos (2Q z)$, we obtain 
\begin{equation}
U_{1111}=U_{2222}=U+J,\,\,\, U_{1221}=U-J, \,\,\,U_{1122}=J
\end{equation}  
where
\begin{align}
U =& \frac{1}{2} \int d\bm{r} d\bm{r}^\prime U(\bm{r}-\bm{r}^\prime)\varphi_0^2(z) \varphi_\perp^2(\bm{\rho})\varphi_0^2(z^\prime)\varphi_\perp^2(\bm{\rho}^\prime) \notag \\
U_{1122} =& \frac{1}{2} \tilde{U}(Q) \int dz\, \varphi_0^4(z) \int d\bm{\rho}\, \varphi_\perp^4(\bm{\rho}) .
\end{align}
Here $\tilde{U}(Q) = 4\pi e^2/(\varepsilon Q^2)$ stands for the Fourier transform of the interaction potential. The direct Coulomb energy can be estimated as $U \sim e^2/(\varepsilon l_\perp)$ and is just the charging energy $E_c$. The exchange energy $U_{1122}$ can be estimated as $U_{1122} \sim e^2/(Q^2 l_zl_\perp^2) \sim U/(Q^2l_z l_\perp) \ll U$.

Therefore, the states with two electrons on the quantum dot can be described by the following Hamiltonian:
\begin{equation}
H_3 =\sum_{\alpha\sigma} \epsilon_\alpha d_{\alpha\sigma}^\dag d_{\alpha\sigma} - J \bm{S}^2 + \lambda_c T^\dag T  \label{ae6}
\end{equation}
where $T = \sum_{\alpha} d_{\alpha \uparrow} d_{\alpha \downarrow}$ and $\lambda_c=J$. The last term in Eq.~\eqref{ae6} describes superconducting correlations. However, since the interaction in the Cooper channel is repulsive ($\lambda_c>0$) one can expect that it will renormalize to zero due to virtual transitions to high energy levels~\cite{ABG}. Then Hamiltonian \eqref{ae3} coincides with Hamiltonian~\eqref{Hqd} projected to the states with two-electrons on the quantum dot.

\section{Explicit expressions for $\chi'$s\label{App2}}

 We present  explicit expressions for the $\chi$'s which one needs to know in order to calculate the co-tunneling current~\eqref{Inin}. Also we present expressions for the averages in Eq.~\eqref{Current4} in terms of the  exact two-particle correlators for the isolated dot ($H_{QD}$) and the Green functions of electrons in the leads which are used to calculate $\chi$'s. We obtain 
\begin{eqnarray}
& & \hspace{-0.5cm} \langle X^\dag(t)H_T(t_1)H_T(t_2)H_T(t_3)\rangle= \notag
\\
& &\hspace{-0.1cm}\langle d^\dag_{\alpha_1t}d_{\alpha_2t_1}d^\dag_{\alpha_3t_2}d_{\alpha_4t_3)}\rangle\,
\tau_{1432}
G^{l>}_{\beta_1\beta_4}(t,t_3)G^{r<}_{\beta_3\beta_2}(t_2,t_1)
\notag\\
&-&\langle d^\dag_{\alpha_1t}d_{\alpha_2t_1}d^\dag_{\alpha_3t_2}d_{\alpha_4t_3}\rangle\,
\tau_{1234}
G^{l>}_{\beta_1\beta_2}(t,t_1)G^{r>}_{\beta_3\beta_4}(t_2,t_3)
\notag\\
&+&\langle d^\dag_{\alpha_1t}d^\dag_{\alpha_2t_1}d_{\alpha_3t_2}d_{\alpha_4t_3}\rangle\,
\tau_{1324}
G^{l>}_{\beta_1\beta_3}(t,t_2)G^{r>}_{\beta_2\beta_4}(t_1,t_3)
\notag\\
&-&\langle d^\dag_{\alpha_1t}d^\dag_{\alpha_2t_1}d_{\alpha_3t_2}d_{\alpha_4t_3}\rangle\,
\tau_{1423}
G^{l>}_{\beta_1\beta_4}(t,t_3)G^{r>}_{\beta_2\beta_3}(t_1,t_2)
\notag\\
&+&\langle d^\dag_{\alpha_1t}d_{\alpha_2t_1}d_{\alpha_3t_2}d^\dag_{\alpha_4t_3}\rangle\,
\tau_{1243}
G^{l>}_{\beta_1\beta_2}(t,t_1)G^{r<}_{\beta_4\beta_3}(t_3,t_2)
\notag\\
&-&\langle d^\dag_{\alpha_1t}d_{\alpha_2t_1}d_{\alpha_3t_2}d^\dag_{\alpha_4t_3}\rangle\,
\tau_{1342}
G^{l>}_{\beta_1\beta_3}(t,t_2)G^{r<}_{\beta_4\beta_2}(t_3,t_1) , \notag\\
&&
\label{C1}
\end{eqnarray}
where $\langle \dots \rangle = \Tr \dots e^{-\beta H_{QD}}/\Tr e^{-\beta H_{QD}}$, $\alpha_k = \{\alpha,\sigma\}$,
$\tau_{ijkl}=\overline{t}^l_{\beta_i\alpha_i}t^l_{\beta_j\alpha_j}\overline{t}^r_{\beta_k\alpha_k}t^r_{\beta_l\alpha_l}$ and
$d_{\alpha t}=d_{\alpha}(t)$. Next 
\begin{eqnarray}
& & \hspace{-0.5cm} \langle H_T(t_2)X^\dag(t)H_T(t_1)H_T(t_3)\rangle= \notag
\\& &\langle d_{\alpha_1t_2}d^\dag_{\alpha_2t}d_{\alpha_3t_1}d^\dag_{\alpha_4t_3}\rangle
\tau_{2341}
G^{l>}_{\beta_2\beta_3}(t,t_1)G^{r<}_{\beta_4\beta_1}(t_3,t_2)
\notag\\
&-&\langle d_{\alpha_1t_2}d^\dag_{\alpha_2t}d_{\alpha_3t_1}d^\dag_{\alpha_4t_3}\rangle
\tau_{2143}
G^{l<}_{\beta_2\beta_1}(t,t_2)G^{r<}_{\beta_4\beta_3}(t_3,t_1)
\notag\\&+&\langle d^\dag_{\alpha_1t_2}d^\dag_{\alpha_2t}d_{\alpha_3t_1}d_{\alpha_4t_3}\rangle
\tau_{2413}
G^{l>}_{\beta_3\beta_4}(t,t_3)G^{r>}_{\beta_1\beta_3}(t_2,t_1)
\notag\\&-&\langle d^\dag_{\alpha_1t_2}d^\dag_{\alpha_2t}d_{\alpha_3t_1}d_{\alpha_4t_3}\rangle
\tau_{2314}
G^{l>}_{\beta_2\beta_3}(t,t_1)G^{r>}_{\beta_1\beta_4}(t_1,t_3)
\notag\\&-&\langle d_{\alpha_1t_2}d^\dag_{\alpha_2t}d^\dag_{\alpha_3t_1}d_{\alpha_4t_3}\rangle
\tau_{2431}
G^{l>}_{\beta_2\beta_4}(t,t_3)G^{r<}_{\beta_3\beta_1}(t_1,t_2)
\notag\\&+&\notag\langle d_{\alpha_1t_2}d^\dag_{\alpha_2t}d^\dag_{\alpha_3t_1}d_{\alpha_4t_3}\rangle
\tau_{2134}
G^{l<}_{\beta_2\beta_1}(t,t_2)G^{r>}_{\beta_3\beta_4}(t_1,t_3) ,\\ && 
\end{eqnarray}

\begin{eqnarray}
& & \hspace{-0.5cm}\langle H_T(t_3)H_T(t_2)X^\dag(t)H_T(t_1)\rangle=\notag
\\& &\langle d_{\alpha_1t_3}d_{\alpha_2t_2}d^\dag_{\alpha_3t}d^\dag_{\alpha_4t_1}\rangle
\tau_{3142}
G^{l<}_{\beta_3\beta_1}(t,t_3)G^{r<}_{\beta_4\beta_2}(t_1,t_2)
\notag\\&-&\langle d_{\alpha_1t_3}d_{\alpha_2t_2}d^\dag_{\alpha_3t}d^\dag_{\alpha_4t_1}\rangle
\tau_{3241}
G^{l<}_{\beta_3\beta_2}(t,t_2)G^{r<}_{\beta_4\beta_1}(t_1,t_3)
\notag\\&-&\langle d^\dag_{\alpha_1t_3}d_{\alpha_2t_2}d^\dag_{\alpha_3t}d_{\alpha_4t_1}\rangle
\tau_{3412}
G^{l>}_{\beta_3\beta_4}(t,t_1)G^{r>}_{\beta_1\beta_2}(t_3,t_2)
\notag\\&+&\langle d^\dag_{\alpha_1t_3}d_{\alpha_2t_2}d^\dag_{\alpha_3t}d_{\alpha_4t_1}\rangle
\tau_{2314}
G^{l>}_{\beta_2\beta_3}(t,t_1)G^{r>}_{\beta_1\beta_4}(t_3,t_2)
\notag\\&-&\langle d_{\alpha_1t_3}d^\dag_{\alpha_2t_2}d^\dag_{\alpha_3t}d_{\alpha_4t_1}\rangle
\tau_{3124}
G^{l<}_{\beta_3\beta_1}(t,t_3)G^{r>}_{\beta_2\beta_4}(t_2,t_1)
\notag\\&+&\notag\langle d_{\alpha_1t_3}d^\dag_{\alpha_2t_2}d^\dag_{\alpha_3t}d_{\alpha_4t_1}\rangle
\tau_{3421}
G^{l>}_{\beta_3\beta_4}(t,t_1)G^{r<}_{\beta_2\beta_1}(t_2,t_3) ,\\ &&
\end{eqnarray}
\begin{eqnarray}
& & \hspace{-0.5cm} \langle H_T(t_1)X^\dag(t)H_T(t_2)H_T(t_3)\rangle=\notag
\\& &\langle d_{\alpha_1t_1}d^\dag_{\alpha_2t}d_{\alpha_3t_2}d^\dag_{\alpha_4t_3}\rangle
\tau_{2341}
G^{l>}_{\beta_2\beta_3}(t,t_2)G^{r<}_{\beta_4\beta_1}(t_3,t_1)
\notag\\&-&\langle d_{\alpha_1t_1}d^\dag_{\alpha_2t}d_{\alpha_3t_2}d^\dag_{\alpha_4t_3}\rangle
\tau_{2143}
G^{l<}_{\beta_2\beta_1}(t,t_1)G^{r<}_{\beta_4\beta_3}(t_3,t_2)
\notag\\&+&\langle d^\dag_{\alpha_1t_1}d^\dag_{\alpha_2t}d_{\alpha_3t_2}d_{\alpha_4t_3}\rangle
\tau_{2413}
G^{l>}_{\beta_2\beta_4}(t,t_3)G^{r>}_{\beta_1\beta_3}(t_1,t_2)
\notag\\&-&\langle d^\dag_{\alpha_1t_1}d^\dag_{\alpha_2t}d_{\alpha_3t_2}d_{\alpha_4t_3}\rangle
\tau_{2314}
G^{l>}_{\beta_2\beta_3}(t,t_2)G^{r>}_{\beta_1\beta_4}(t_1,t_3)
\notag\\&-&\langle d_{\alpha_1t_1}d^\dag_{\alpha_2t}d^\dag_{\alpha_3t_2}d_{\alpha_4t_3}\rangle
\tau_{2431}
G^{l>}_{\beta_2\beta_4}(t,t_3)G^{r<}_{\beta_3\beta_1}(t_2,t_1)
\notag\\&+&\notag\langle d_{\alpha_1t_1}d^\dag_{\alpha_2t}d^\dag_{\alpha_3t_2}d_{\alpha_4t_3}\rangle
\tau_{2134}
G^{l<}_{\beta_2\beta_1}(t,t_1)G^{r>}_{\beta_3\beta_4}(t_2,t_3) ,\\ &&
\end{eqnarray}
\begin{eqnarray}
& & \hspace{-0.5cm} \langle H_T(t_3)H_T(t_1)X^\dag(t)H_T(t_2))\rangle
=\notag\\& &\langle d_{\alpha_1t_3}d_{\alpha_2t_1}d^\dag_{\alpha_3t}d^\dag_{\alpha_4t_2}\rangle
\tau_{3142}
G^{l<}_{\beta_3\beta_1}(t,t_3)G^{r<}_{\beta_4\beta_2}(t_2,t_1)
\notag\\&-&\langle d_{\alpha_1t_3}d_{\alpha_2t_1}d^\dag_{\alpha_3t}d^\dag_{\alpha_4t_2}\rangle
\tau_{3241}
G^{l<}_{\beta_3\beta_2}(t,t_1)G^{r<}_{\beta_4\beta_1}(t_2,t_3)
\notag\\&-&\langle d^\dag_{\alpha_1t_3}d_{\alpha_2t_1}d^\dag_{\alpha_3t}d_{\alpha_4t_2}\rangle
\tau_{3412}
G^{l>}_{\beta_3\beta_4}(t,t_2)G^{r>}_{\beta_1\beta_2}(t_3,t_1)
\notag\\&+&\langle d^\dag_{\alpha_1t_3}d_{\alpha_2t_1}d^\dag_{\alpha_3t}d_{\alpha_4t_2}\rangle
\tau_{3214}
G^{l<}_{\beta_3\beta_2}(t,t_1)G^{r>}_{\beta_1\beta_4}(t_3,t_2)
\notag\\&-&\langle d_{\alpha_1t_3}d^\dag_{\alpha_2t_1}d^\dag_{\alpha_3t}d_{\alpha_4t_2}\rangle
\tau_{3124}
G^{l<}_{\beta_3\beta_1}(t,t_3)G^{r>}_{\beta_2\beta_4}(t_1,t_2)
\notag\\&+&\langle d_{\alpha_1t_1}d^\dag_{\alpha_2t}d_{\alpha_3t_2}d^\dag_{\alpha_4t_3}\rangle
\tau_{2341}
G^{l>}_{\beta_2\beta_3}(t,t_2)G^{r<}_{\beta_4\beta_1}(t_3,t_1)\notag 
\\&+&\notag
\langle d_{\alpha_1t_3}d^\dag_{\alpha_2t_1}d^\dag_{\alpha_3t}d_{\alpha_4t_2}\rangle
\tau_{3421}
G^{l>}_{\beta_3\beta_4}(t,t_2)G^{r<}_{\beta_2\beta_1}(t_1,t_3) ,\\ &&
\end{eqnarray}
\begin{eqnarray}
& & \hspace{-0.5cm} \langle H_T(t_2)H_T(t_1)X^\dag(t)H_T(t_3)\rangle=\notag
\\& &\langle d_{\alpha_1t_2}d_{\alpha_2t_1}d^\dag_{\alpha_3t}d^\dag_{\alpha_4t_3}\rangle
\tau_{3142}
G^{l<}_{\beta_3\beta_1}(t,t_2)G^{r<}_{\beta_4\beta_2}(t_3,t_1)
\notag\\&-&\langle d_{\alpha_1t_2}d_{\alpha_2t_1}d^\dag_{\alpha_3t}d^\dag_{\alpha_4t_3}\rangle
\tau_{3241}
G^{l<}_{\beta_3\beta_2}(t,t_1)G^{r<}_{\beta_4\beta_1}(t_3,t_2)
\notag\\&-&\langle d^\dag_{\alpha_1t_2}d_{\alpha_2t_1}d^\dag_{\alpha_3t}d_{\alpha_4t_3}\rangle
\tau_{3412}
G^{l>}_{\beta_3\beta_4}(t,t_3)G^{r>}_{\beta_1\beta_2}(t_2,t_1)
\notag\\&+&\langle d^\dag_{\alpha_1t_2}d_{\alpha_2t_1}d^\dag_{\alpha_3t}d_{\alpha_4t_3}\rangle
\tau_{3214}
G^{l<}_{\beta_3\beta_2}(t,t_1)G^{r>}_{\beta_1\beta_4}(t_2,t_3)
\notag\\&-&\langle d_{\alpha_1t_2}d^\dag_{\alpha_2t_1}d^\dag_{\alpha_3t}d_{\alpha_4t_3}\rangle
\tau_{3124}
G^{l<}_{\beta_3\beta_1}(t,t_2)G^{r>}_{\beta_2\beta_4}(t_1,t_3)
\notag\\&+&\notag\langle d_{\alpha_1t_2}d^\dag_{\alpha_2t_1}d^\dag_{\alpha_3t}d_{\alpha_4t_3}\rangle
\tau_{3421}
G^{l>}_{\beta_3\beta_4}(t,t_3)G^{r<}_{\beta_2\beta_1}(t_1,t_2) ,\\ &&
\end{eqnarray}
\begin{eqnarray}
& & \hspace{-0.5cm} \langle H_T(t_3)H_T(t_2)H_T(t_1)X^\dag(t)\rangle=\notag
\\& &\langle d_{\alpha_1t_3}d_{\alpha_2t_2}d^\dag_{\alpha_3t_1}d^\dag_{\alpha_4t}\rangle
\tau_{4231}
G^{l<}_{\beta_4\beta_2}(t,t_2)G^{r<}_{\beta_3\beta_1}(t_1,t_3)
\notag\\&-&\langle d_{\alpha_1t_3}d_{\alpha_2t_2}d^\dag_{\alpha_3t_1}d^\dag_{\alpha_4t}\rangle
\tau_{4132}
G^{l<}_{\beta_4\beta_1}(t,t_3)G^{r<}_{\beta_3\beta_2}(t_1,t_2)
\notag\\&-&\langle d_{\alpha_1t_3}d^\dag_{\alpha_2t_2}d_{\alpha_3t_1}d^\dag_{\alpha_4t}\rangle
\tau_{4321}
G^{l<}_{\beta_4\beta_3}(t,t_1)G^{r<}_{\beta_2\beta_1}(t_2,t_3)
\notag\\&+&\langle d_{\alpha_1t_3}d^\dag_{\alpha_2t_2}d_{\alpha_3t_1}d^\dag_{\alpha_4t}\rangle
\tau_{4123}
G^{l<}_{\beta_4\beta_1}(t,t_3)G^{r>}_{\beta_2\beta_3}(t_2,t_1)
\notag\\&+&\langle d^\dag_{\alpha_1t_3}d_{\alpha_2t_2}d_{\alpha_3t_1}d^\dag_{\alpha_4t}\rangle
\tau_{4312}
G^{l<}_{\beta_4\beta_3}(t,t_1)G^{r>}_{\beta_1\beta_2}(t_3,t_2)
\notag\\&-&\notag\langle d^\dag_{\alpha_1t_3}d_{\alpha_2t_2}d_{\alpha_3t_1}d^\dag_{\alpha_4t}\rangle
\tau_{4213}
G^{l<}_{\beta_4\beta_2}(t,t_2)G^{r>}_{\beta_1\beta_3}(t_3,t_1) .\\ &&\label{C8}
\end{eqnarray}
\begin{eqnarray}
& & \hspace{-0.5cm} \langle H_T(t_3)X^\dag(t)H_T(t_1)H_T(t_2)\rangle=\notag
\\& &\langle d_{\alpha_1t_3}d^\dag_{\alpha_2t}d_{\alpha_3t_1}d^\dag_{\alpha_4t_2}\rangle
\tau_{2341}
G^{l>}_{\beta_2\beta_3}(t,t_2)G^{r<}_{\beta_4\beta_1}(t_3,t_1)
\notag\\&-&\langle d_{\alpha_1t_3}d^\dag_{\alpha_2t}d_{\alpha_3t_1}d^\dag_{\alpha_4t_2}\rangle
\tau_{2143}
G^{l<}_{\beta_2\beta_1}(t,t_3)G^{r<}_{\beta_4\beta_3}(t_3,t_1)
\notag\\&+&\langle d^\dag_{\alpha_1t_3}d^\dag_{\alpha_2t}d_{\alpha_3t_1}d_{\alpha_4t_2}\rangle
\tau_{2413}
G^{l>}_{\beta_2\beta_4}(t,t_2)G^{r>}_{\beta_1\beta_3}(t_3,t_1)
\notag\\&-&\langle d^\dag_{\alpha_1t_3}d^\dag_{\alpha_2t}d_{\alpha_3t_1}d_{\alpha_4t_2}\rangle
\tau_{2314}
G^{l>}_{\beta_2\beta_3}(t,t_1)G^{r>}_{\beta_1\beta_4}(t_3,t_2)
\notag\\&-&\langle d_{\alpha_1t_3}d^\dag_{\alpha_2t}d^\dag_{\alpha_3t_1}d_{\alpha_4t_2}\rangle
\tau_{2431}
G^{l>}_{\beta_2\beta_4}(t,t_2)G^{r<}_{\beta_3\beta_1}(t_1,t_3)
\notag\\ \notag&+&\langle d_{\alpha_1t_3}d^\dag_{\alpha_2t}d^\dag_{\alpha_3t_1}d_{\alpha_4t_2}\rangle
\tau_{2134}
G^{l<}_{\beta_2\beta_1}(t,t_3)G^{r>}_{\beta_3\beta_4}(t_1,t_2) ,\\ &&
\end {eqnarray}
The non-interference contributions to $\chi$'s are given as
\begin{eqnarray}
\chi^{<,>}&\equiv&\Delta^2 \Re \int_{-\infty}^t dt_1\int_{-\infty}^{t_1} dt_2\int_{-\infty}^{t_2} dt_3\Bigl (\notag\\
&+&g^l_{\alpha_1}g^r_{\alpha_3}\langle d_{\alpha_1t_3}d^\dag_{\alpha_1t}d^\dag_{\alpha_3t_1}d_{\alpha_3t_2}\rangle
\notag\\&+&g^l_{\alpha_1}g^r_{\alpha_3}\langle d_{\alpha_1t_2}d^\dag_{\alpha_1t}d^\dag_{\alpha_3t_1}d_{\alpha_3t_3}\rangle
\notag\\&-&g^l_{\alpha_1}g^r_{\alpha_2}\langle d_{\alpha_1t_3}d^\dag_{\alpha_2t_2}d^\dag_{\alpha_1t}d_{\alpha_2t_1}\rangle
\notag\\&+&g^l_{\alpha_1}g^r_{\alpha_3}\langle d_{\alpha_1t_1}d^\dag_{\alpha_1t}d^\dag_{\alpha_3t_2}d_{\alpha_3t_3}\rangle
\notag\\&+&g^l_{\alpha_2}g^r_{\alpha_1}\langle d^\dag_{\alpha_1t_3}d_{\alpha_2t_1}d^\dag_{\alpha_2t}d_{\alpha_1t_2}\rangle
\notag\\&+&g^l_{\alpha_3}g^r_{\alpha_1}\langle d_{\alpha_1t_3}d^\dag_{\alpha_1t_1}d^\dag_{\alpha_3t}d_{\alpha_3t_2}\rangle
\notag\\&+&g^l_{\alpha_2}g^r_{\alpha_1}\langle d^\dag_{\alpha_1t_2}d_{\alpha_2t_1}d^\dag_{\alpha_2t}d_{\alpha_1t_3}\rangle
\notag\\&-&g^l_{\alpha_1}g^r_{\alpha_2}\langle d_{\alpha_1t_2}d^\dag_{\alpha_2t_1}d^\dag_{\alpha_1t}d_{\alpha_2t_3}\rangle
\notag\\&+&g^l_{\alpha_1}g^r_{\alpha_2}\langle d_{\alpha_1t_3}d^\dag_{\alpha_2t_2}d_{\alpha_2t_1}d^\dag_{\alpha_1t}\rangle
\notag\\&+&g^l_{\alpha_3}g^r_{\alpha_1}\langle d^\dag_{\alpha_1t_3}d_{\alpha_1t_2}d_{\alpha_3t_1}d^\dag_{\alpha_3t}\rangle
\notag\\&-&g^l_{\alpha_2}g^r_{\alpha_1}\langle d^\dag_{\alpha_1t_3}d_{\alpha_2t_2}d_{\alpha_1t_1}d^\dag_{\alpha_2t}\rangle \Bigr ),\label{ch1}
\end{eqnarray}
\begin{eqnarray}
\chi^{>,<}\equiv\Delta^2 \Re\int_{-\infty}^t dt_1\int_{-\infty}^{t_1} dt_2\int_{-\infty}^{t_2} dt_3\Bigl(
\notag\\+g^l_{\alpha_1}g^r_{\alpha_2}\langle d^\dag_{\alpha_1t}d_{\alpha_2t_1}d^\dag_{\alpha_2t_2}d_{\alpha_1t_3}\rangle
\notag\\+g^l_{\alpha_1}g^r_{\alpha_3}\langle d^\dag_{\alpha_1t}d_{\alpha_1t_1}d_{\alpha_3t_2}d^\dag_{\alpha_3t_3}\rangle
\notag\\-g^l_{\alpha_1}g^r_{\alpha_2}\langle d^\dag_{\alpha_1t}d_{\alpha_2t_1}d_{\alpha_1t_2}d^\dag_{\alpha_1t_3}\rangle
\notag\\+g^l_{\alpha_2}g^r_{\alpha_1}\langle d_{\alpha_1t_3}d^\dag_{\alpha_2t}d_{\alpha_2t_1}d^\dag_{\alpha_1t_2}\rangle
\notag\\-g^l_{\alpha_2}g^r_{\alpha_1}\langle d_{\alpha_1t_3}d^\dag_{\alpha_2t}d^\dag_{\alpha_1t_1}d_{\alpha_2t_2}\rangle
\notag\\+g^l_{\alpha_2}g^r_{\alpha_1}\langle d_{\alpha_1t_2}d^\dag_{\alpha_2t}d_{\alpha_2t_1}d^\dag_{\alpha_1t_3}\rangle
\notag\\-g^l_{\alpha_2}g^r_{\alpha_1}\langle d_{\alpha_1t_2}d^\dag_{\alpha_2t}d^\dag_{\alpha_1t_1}d_{\alpha_2t_3}\rangle
\notag\\+g^l_{\alpha_3}g^r_{\alpha_1}\langle d_{\alpha_1t_3}d^\dag_{\alpha_1t_2}d^\dag_{\alpha_3t}d_{\alpha_3t_1}\rangle
\notag\\+g^l_{\alpha_2}g^r_{\alpha_1}\langle d_{\alpha_1t_1}d^\dag_{\alpha_2t}d_{\alpha_2t_2}d^\dag_{\alpha_1t_3}\rangle
\notag\\-g^l_{\alpha_2}g^r_{\alpha_1}\langle d_{\alpha_1t_1}d^\dag_{\alpha_2t}d^\dag_{\alpha_1t_2}d_{\alpha_2t_3}\rangle
\notag\\+g^l_{\alpha_3}g^r_{\alpha_1}\langle d_{\alpha_1t_3}d^\dag_{\alpha_1t_1}d^\dag_{\alpha_3t}d_{\alpha_3t_2}\rangle
\notag\\+g^l_{\alpha_3}g^r_{\alpha_1}\langle d_{\alpha_1t_2}d^\dag_{\alpha_1t_1}d^\dag_{\alpha_3t}d_{\alpha_3t_3}\rangle\Bigr),\label{ch2}
\end{eqnarray}
\begin{eqnarray}
\chi^{>,>}\equiv\Delta^2 \Re\int_{-\infty}^t dt_1\int_{-\infty}^{t_1} dt_2\int_{-\infty}^{t_2} dt_3\Bigl(
\notag\\-g^l_{\alpha_1}g^r_{\alpha_3}\langle d^\dag_{\alpha_1t}d_{\alpha_1t_1}d^\dag_{\alpha_3t_2}d_{\alpha_3t_3}\rangle
\notag\\+g^l_{\alpha_1}g^r_{\alpha_2}\langle d^\dag_{\alpha_1t}d^\dag_{\alpha_2t_1}d_{\alpha_1t_2}d_{\alpha_2t_3}\rangle
\notag\\-g^l_{\alpha_1}g^r_{\alpha_2}\langle d^\dag_{\alpha_1t}d^\dag_{\alpha_2t_1}d_{\alpha_2t_2}d_{\alpha_1t_3}\rangle
\notag\\+g^l_{\alpha_2}g^r_{\alpha_1}\langle d^\dag_{\alpha_1t_3}d^\dag_{\alpha_2t}d_{\alpha_1t_1}d_{\alpha_2t_2}\rangle
\notag\\-g^l_{\alpha_2}g^r_{\alpha_1}\langle d^\dag_{\alpha_1t_3}d^\dag_{\alpha_2t}d_{\alpha_2t_1}d_{\alpha_1t_2}\rangle
\notag\\+g^l_{\alpha_2}g^r_{\alpha_1}\langle d^\dag_{\alpha_1t_2}d^\dag_{\alpha_2t}d_{\alpha_1t_1}d_{\alpha_2t_3}\rangle
\notag\\-g^l_{\alpha_2}g^r_{\alpha_1}\langle d^\dag_{\alpha_1t_2}d^\dag_{\alpha_2t}d_{\alpha_2t_1}d_{\alpha_1t_3}\rangle
\notag\\-g^l_{\alpha_3}g^r_{\alpha_1}\langle d^\dag_{\alpha_1t_3}d_{\alpha_1t_2}d^\dag_{\alpha_3t}d_{\alpha_3t_1}\rangle
\notag\\+g^l_{\alpha_2}g^r_{\alpha_1}\langle d^\dag_{\alpha_1t_3}d_{\alpha_2t_2}d^\dag_{\alpha_2t}d_{\alpha_1t_1}\rangle
\notag\\+g^l_{\alpha_2}g^r_{\alpha_1}\langle d^\dag_{\alpha_1t_1}d^\dag_{\alpha_2t}d_{\alpha_1t_2}d_{\alpha_2t_3}\rangle
\notag\\-g^l_{\alpha_2}g^r_{\alpha_1}\langle d^\dag_{\alpha_1t_1}d^\dag_{\alpha_2t}d_{\alpha_2t_2}d_{\alpha_1t_3}\rangle
\notag\\-g^l_{\alpha_3}g^r_{\alpha_1}\langle d^\dag_{\alpha_1t_3}d_{\alpha_1t_1}d^\dag_{\alpha_3t}d_{\alpha_3t_2}\rangle
\notag\\-g^l_{\alpha_3}g^r_{\alpha_1}\langle d^\dag_{\alpha_1t_2}d_{\alpha_1t_1}d^\dag_{\alpha_3t}d_{\alpha_3t_3}\rangle\Bigr),\label{ch3}
\end{eqnarray}
\begin{eqnarray}
\chi^{<,<}\equiv\Delta^2 \Re\int_{-\infty}^t dt_1\int_{-\infty}^{t_1} dt_2\int_{-\infty}^{t_2} dt_3\Bigl(\notag\\-g^l_{\alpha_1}g^r_{\alpha_3}\langle d_{\alpha_1t_3}d^\dag_{\alpha_1t}d_{\alpha_3t_1}d^\dag_{\alpha_3t_2}\rangle
\notag\\-g^l_{\alpha_1}g^r_{\alpha_3}\langle d_{\alpha_1t_2}d^\dag_{\alpha_1t}d_{\alpha_3t_1}d^\dag_{\alpha_3t_3}\rangle
\notag\\+g^l_{\alpha_1}g^r_{\alpha_2}\langle d_{\alpha_1t_3}d_{\alpha_2t_2}d^\dag_{\alpha_1t}d^\dag_{\alpha_2t_1}\rangle
\notag\\-g^l_{\alpha_2}g^r_{\alpha_1}\langle d_{\alpha_1t_3}d_{\alpha_2t_2}d^\dag_{\alpha_2t}d^\dag_{\alpha_1t_1}\rangle
\notag\\-g^l_{\alpha_1}g^r_{\alpha_3}\langle d_{\alpha_1t_1}d^\dag_{\alpha_1t}d_{\alpha_3t_2}d^\dag_{\alpha_3t_3}\rangle
\notag\\+g^l_{\alpha_1}g^r_{\alpha_2}\langle d_{\alpha_1t_3}d_{\alpha_2t_1}d^\dag_{\alpha_1t}d^\dag_{\alpha_2t_2}\rangle
\notag\\-g^l_{\alpha_2}g^r_{\alpha_1}\langle d_{\alpha_1t_3}d_{\alpha_2t_1}d^\dag_{\alpha_2t}d^\dag_{\alpha_1t_2}\rangle
\notag\\+g^l_{\alpha_1}g^r_{\alpha_2}\langle d_{\alpha_1t_2}d_{\alpha_2t_1}d^\dag_{\alpha_1t}d^\dag_{\alpha_2t_3}\rangle
\notag\\-g^l_{\alpha_2}g^r_{\alpha_1}\langle d_{\alpha_1t_2}d_{\alpha_2t_1}d^\dag_{\alpha_2t}d^\dag_{\alpha_1t_3}\rangle
\notag\\+g^l_{\alpha_2}g^r_{\alpha_1}\langle d_{\alpha_1t_3}d_{\alpha_2t_2}d^\dag_{\alpha_1t_1}d^\dag_{\alpha_2t}\rangle
\notag\\-g^l_{\alpha_1}g^r_{\alpha_2}\langle d_{\alpha_1t_3}d_{\alpha_2t_2}d^\dag_{\alpha_2t_1}d^\dag_{\alpha_1t}\rangle
\notag\\-g^l_{\alpha_3}g^r_{\alpha_1}\langle d_{\alpha_1t_3}d^\dag_{\alpha_1t_2}d_{\alpha_3t_1}d^\dag_{\alpha_3t}\rangle\Bigr).\label{ch4}
\end{eqnarray}

\end{document}